\documentclass[prx,twocolumn,showpacs,amsmath,amssymb,floatfix,superscriptaddress]{revtex4}
\usepackage{color,graphics,epsfig,rotating}
\usepackage{graphicx}
\usepackage{dcolumn}
\usepackage{bm}
\usepackage{subfigure}
\usepackage{mathrsfs}
\usepackage{lmodern}

\usepackage{grffile}
\usepackage{xspace}

\def\beq{\begin{equation}}
\def\eeq{\end{equation}}

\newcommand{\lunio}{LuNiO$_{3}$\xspace}

\newcommand{\ortho}{orthorhombic\xspace}
\newcommand{\mono}{monoclinic\xspace}

\usepackage[svgnames]{xcolor}
\newif\ifshowcomments\showcommentstrue

\newcommand{\RNO}{\textsl{R}NiO$_3$\,}
\newcommand{\LNO}{LaNiO$_3$\,}
\newcommand{\LuNO}{LuNiO$_3$\,}
\newcommand{\Tmit}{T_{\mathrm{MIT}}}
\newcommand{\Tmag}{T_{\mathrm{N}}}
\newcommand{\eg}{e_{g}}
\newcommand{\ttg}{t_{2g}}
\newcommand\Lbar{\underline{L}}
\newcommand{\delsite}{\Delta_s\,}
\newcommand{\delc}{\Delta_s^c\,}
\def\vk{{\bf k}}
\def\hn{\hat{n}}

\def\mp{m^\prime}
\def\hnmu{\hat{n}_{m\uparrow}}
\def\hnmd{\hat{n}_{m\downarrow}}

\def\hnpmd{\hat{n}_{m^\prime\downarrow}}
\def\spinup{\uparrow}
\def\spindown{\downarrow}
\newcommand{\Hint}{H_{\mathrm{int}}}

\newcommand{\Wlarge}{W_{>}}
\newcommand{\Wsmall}{W_{<}}

\newcommand{\LBlow}{$\mathrm{LB}_{<}\,$}
\newcommand{\LBhi}{$\mathrm{LB}_{>}\,$}
\newcommand{\SBlow}{$\mathrm{SB}_{<}\,$}
\newcommand{\SBhi}{$\mathrm{SB}_{>}\,$}

\newcommand{\teff}{t_{\mathrm{eff}}}
\newcommand{\Weff}{W_{\mathrm{eff}}}
\newcommand{\Ueff}{U_{\mathrm{eff}}}
\newcommand{\Ucs}{U_c^{<}}
\newcommand{\Ucl}{U_c^{>}}

\begin{document}

\title{A low-energy description of the metal-insulator transition in
  the rare-earth nickelates}

\author{Alaska Subedi}
\affiliation{Max Planck Institute for the Structure and Dynamics of
  Matter, 22761 Hamburg, Germany}
\affiliation{Centre de Physique Th\'eorique, Ecole Polytechnique,
  CNRS, 91128 Palaiseau Cedex, France}
\author{Oleg E. Peil}
\affiliation{DPMC, Universit\'e de Gen\`eve, CH-1211 Gen\`eve, Switzerland}
\affiliation{Centre de Physique Th\'eorique, Ecole Polytechnique, CNRS, 91128 Palaiseau Cedex, France}
\author{Antoine Georges}
\affiliation{Coll\`ege de France, 11 place Marcelin Berthelot, 75005 Paris, France}
\affiliation{Centre de Physique Th\'eorique, Ecole Polytechnique, CNRS, 91128 Palaiseau Cedex, France}
\affiliation{DPMC, Universit\'e de Gen\`eve, CH-1211 Gen\`eve, Switzerland}

\date{\today}

\begin{abstract}
We propose a simple theoretical description of the metal-insulator
transition of rare-earth nickelates. The theory involves only two
orbitals per nickel site, corresponding to the low-energy anti-bonding
$e_g$ states. In the monoclinic insulating state, bond-length
disproportionation splits the manifold of $e_g$ bands, corresponding
to a modulation of the effective on-site energy.  We show that, when
subject to a local Coulomb repulsion $U$ and Hund's coupling $J$, the
resulting bond-disproportionated state is a paramagnetic insulator for
a wide range of interaction parameters. Furthermore, we find that when
$U-3J$ is small or negative, a spontaneous instability to bond
disproportionation takes place for large enough $J$. This minimal
theory emphasizes that a small or negative charge-transfer energy, a
large Hund's coupling, and a strong coupling to
bond-disproportionation are the key factors underlying the
transition. Experimental consequences of this theoretical picture are
discussed.
\end{abstract}

\maketitle

\section{Introduction 
}
\label{sec:intro}

The rare-earth nickelate series \RNO displays a rich phase diagram
with striking structural and metal-insulator
transitions~\cite{Torrance1992} (see Refs.~\onlinecite{Medarde1997}
and \onlinecite{Catalan2008} for reviews).  Only the end member of the
series \LNO is metallic and non-magnetic down to the lowest
temperature~\cite{Goodenough1965}. Compounds with heavier rare-earth
ions \textsl{R} (\textit{i.e.}\ smaller ionic radius) display a
bad-metal~\cite{Jaramillo2014} to insulator transition (MIT) as
temperature is lowered. The compounds with Pr or Nd undergo a
transition to a magnetic insulating state. However, the transition is
towards a non-magnetic insulating state for the smaller rare-earth
ions (\textsl{R} $=$ Sm and beyond), and the magnetic ordering sets in
only at a lower temperature. For example, in \LuNO the MIT occurs at
$\Tmit\simeq 600$~K, but the magnetic ordering occurs only at
$\Tmag\simeq 130$~K.

The transition from the high-temperature metallic phase in the
orthorhombic ($Pbnm$) structure to the low temperature insulating
phase is concomitant with a structural transition to a monoclinic
(P2$_1/n$) structure with two types of NiO$_6$
octahedra~\cite{Alonso1999,Alonso2000,Alonso2001,Piamonteze2005}.
One set of octahedra is compressed and has short Ni-O bonds (SB), and
the other set of octahedra is expanded and has long Ni-O bonds (LB).

The nature of this MIT has still not been fully clarified, and the
respective role of correlations (on-site Coulomb repulsion $U$ and
Hund's rule coupling $J$), orbital degeneracy, and structural
transition in causing the MIT is still under debate.  Interest in this
issue has been renewed by the recent activity on nickelates in the
form of thin films and heterostructures, and the opportunities for
controlling the MIT in these structures (e.g. by strain or electric
field)~\cite{Scherwitzl2009,Scherwitzl2010,Scherwitzl2011,Zubko2011,Rondinelli2011,Hwang2012,Disa2013,Liu2013}.
The broad question behind the nature of the MIT is how we should think
of the electronic structure of these materials.

In the most naive ionic picture, each nickel is Ni$^{3+}$ ($d^7$)
corresponding to the low-spin configuration $\ttg^6\eg^1$.  The MIT
was initially interpreted as a Mott transition (or more accurately a
metal to charge-transfer insulator transition) of the quarter-filled
$\eg$ shell caused by the change in Ni-O-Ni angle as the tolerance
factor is reduced~\cite{Torrance1992}.
However, the MIT is simultaneously accompanied by the structural
differentiation between the two nickel sites
and there is no anomaly in the susceptibility at
$\Tmit$~\cite{Zhou2000}, which makes the homogeneous Mott transition picture 
untenable (see also Ref.~\onlinecite{Wang2012}).
Clearly, the MIT is not a Slater transition either, since for smaller rare-earths compounds 
it occurs at a temperature above that of magnetic ordering. 


Although $\ttg^6\eg^1$ is an orbitally degenerate configuration that
is susceptible to Jahn-Teller distortion of the octahedra, such a
distortion is not experimentally observed for all
octahedra~\cite{Zhou_bonding_2004}. Nonetheless, several experiments
reveal that the lattice degrees of freedom play an active role in the
transition.  For example, a large isotope effect is observed on
$\Tmit$, at least for the lighter rare
earths~\cite{Medarde_prl_1998,Medarde_physicaB_1998}. In addition,
recent control of the transition by light pulses resonant with 
specific vibrational modes have emphasized the importance of lattice
degrees of freedom~\cite{Caviglia2012}. This suggests that these
materials take some other structural route to lift the orbital
degeneracy.

Charge disproportionation on the Ni sites into Ni$^{3+\delta}$ (on SB sites) and
Ni$^{3-\delta}$ (on LB sites) has been proposed as an alternative way to
interpret the structural transition and resolve the orbital degeneracy issue.
%
%
A number of recent experiments, especially resonant spectroscopies
\cite{Alonso1999,Alonso2001,Staub2002,Caytuero2007,Medarde2008,Bodenthin2011}
have been interpreted in terms of charge
disproportionation. However, a strong on-site $d$$-$$d$ Coulomb
repulsion $U$ that is likely relevant for Ni should suppress charge
disproportionation.

A different picture was proposed early on by Demourgues \textit{et
  al.}~\cite{Demourgues1993}, and Mizokawa, Khomskii and
Sawatzky~\cite{Mizokawa2000}.  The basic premise of these authors is
that the $d^7$ ionic assignment is invalid: holes are formed on oxygen
sites, a behaviour sometimes referred to as `negative charge-transfer'
insulators~\cite{Mizokawa1991,Mizokawa1994,Ushakov2011}.
Recent work by Park, Millis and
Marianetti~\cite{Park2012,Park2014,Lau2013} provides strong
theoretical support to the importance of ligand holes. These authors
performed electronic structure calculations involving both oxygen and
nickel states, taking correlation effects into account within
dynamical mean-field theory (DMFT).  They found that the insulating
phase can be viewed as a `site-selective' Mott phase.
The extreme limit of this picture is the following~\cite{Park2012}, as
also emphasized by Johnston {\it et al.}~\cite{Johnston2014}: LB
octahedra have the configuration $d^8$, while SB octahedra acquire the
configuration $d^8\Lbar^2$, with two holes on the ligand and two
electrons in the atomic-like $\eg$ shell of nickel, hence lifting
orbital degeneracy.
Note that this does not imply an actual charge disproportionation
since each oxygen is shared by two octahedra, although the charge
density is larger in the SB octahedra than in the LB octahedra.
%
%
%
The spin is strongly modulated in this picture, with $S=0$ on the SB
octahedra (where the two ligand holes screen out the Ni moment) and
$S=1$ on the LB octahedra. In the strong coupling limit, the system can be
described as the superposition of a Kondo insulator on the SB sites
and of a Mott insulator on the LB sites, a mechanism called
`site-selective Mott transition' by the authors of
Ref.~\cite{Park2012}.
This mechanism is also qualitatively consistent with ideas of
Goodenough and coworkers pointing at the strong covalent bonding in
the SB octahedra~\cite{Zhou_bonding_2004,Goodenough1996,Catalan2008}.
This picture is quite attractive and reconciles a number of different
experimental observations, in particular the absence of Jahn-Teller
distortions. 
%
%

Mazin \textit{et al.} have pointed out that a disproportionation of
the type $2\eg^1 \rightarrow \eg^0 + \eg^2$ can be favorable when $U -
3J$ is small \cite{Mazin2007}. This picture is similar to that
mentioned above if one assumes that the disproportionation occurs for
the antibonding $\eg$ orbital resulting from the strong hybridization
between Ni $3d$ and O $2p$ states. Ref.~\cite{Mazin2007} used
density functional calculations to show that magnetically ordered
state with such disproportionation is indeed an insulator, although this
picture cannot fully describe the non-magnetic insulating state that
occurs for the majority of the rare-earth ions.

Two outstanding theoretical questions remain unanswered to this day
however.  The first one is whether it is at all possible to construct
a low-energy description of nickelates and of their MIT in terms of
{\it low-energy electronic states only}.  By low energy, we mean a
two-band (per nickel site) picture involving only $\eg$ states
resulting from the strong hybridization between oxygen and nickel
atomic states.
This question was previously addressed by Lee, Chen and Balents in the
weak-coupling limit, mostly in connection with ordered
states~\cite{Lee2011_prl,Lee2011_prb}.
The second question is why the metallic state of these materials is so
easily destabilized into an insulator with bond-length
disproportionation and lower crystal symmetry. These are the questions
that we set out to answer in the present paper.

Our answer to the first question is in fact remarkably simple. We find
that an effective two-band description is indeed possible, provided that
the effective low-energy interaction $U-3J$ between two electrons with
parallel spin in different orbitals is smaller than the energy
difference $\delsite$ between inequivalent nickel sites.  Since
$\delsite$ is zero in the orthorombic phase and remains a small energy
scale in the monoclinic phase (our estimate for \LuNO is
$\delsite\simeq 0.25$~eV), this implies that $U-3J$ has to be taken
small or even slightly negative, in qualitative agreement with the
negative charge-transfer picture.  We emphasize that in this
description $J$ and, especially, $U$ are not the values for
atomic-like localized nickel states but rather renormalized low-energy
values appropriate for the covalent $\eg$ states.
We explicitly construct such a low-energy model by performing
electronic structure calculations of \LuNO in both the
high-temperature orthorhombic and low-temperature monoclinic
structures.  By exploring the phase diagram of this model as a
function of $U$ and $J$ using DMFT, we show that a consistent
description of both the metallic and the insulating phases can be
obtained.

Furthermore, our low-energy description also provides an insight into the
second question.  We show that the phase-diagram of this model as a
function of $U$ and $\delsite$ changes drastically as the Hund's coupling $J$ is increased. 
When $U-3J$ is small ($\leq\delsite$), a symmetry breaking transition of the metal into 
a spontaneously disproportionated insulating state takes place. 
This confirms the importance of Hund's coupling for these
materials~\cite{Mazin2007} and provides a new low-energy perspective
on its physical relevance.

Our results also clearly establish that the homogeneous quarter-filled
Mott transition scenario is untenable and that the MIT is a
cooperative effect between the electronic degrees of freedom and the
lattice distortion, which plays an essential role.  Our description
has direct implications for experiments probing excitations in both
phases, such as optical spectroscopy, as discussed at the end of the
paper.


This article is organized as follows. In Sec.~\ref{sec:elstruc} we
introduce the low-energy model and discuss the electronic structure of
both the high-$T$ orthorombic and low-$T$ monoclinic phases of
\LuNO. In Sec.~\ref{sec:phases}, we then explore the phase diagram of
this model using DMFT, for both phases, as a function of $U$ and $J$
and identify the region of interaction parameters which is appropriate
to the description of nickelates.  In Sec.~\ref{sec:discussion}, we
provide a qualitative understanding of the physics of the problem, by
considering a simplified model in which $\delsite$ can be varied
continuously, and demonstrate the sensitivity to
site-disproportionation when the Hund's coupling $J$ is large enough.
Finally, in Sec.~\ref{sec:exp} we discuss
consequences for experiments such as photoemission,
optical spectroscopy and magnetic probes such as NMR, focusing on
relevant physical observables.

\section{Electronic structure and low-energy model}
\label{sec:elstruc}

\subsection{Electronic structure of LuNiO$_3$}
\label{sec:elstruc_bands}


\begin{figure}
\begin{tabular}{c}
\includegraphics[width=\columnwidth]{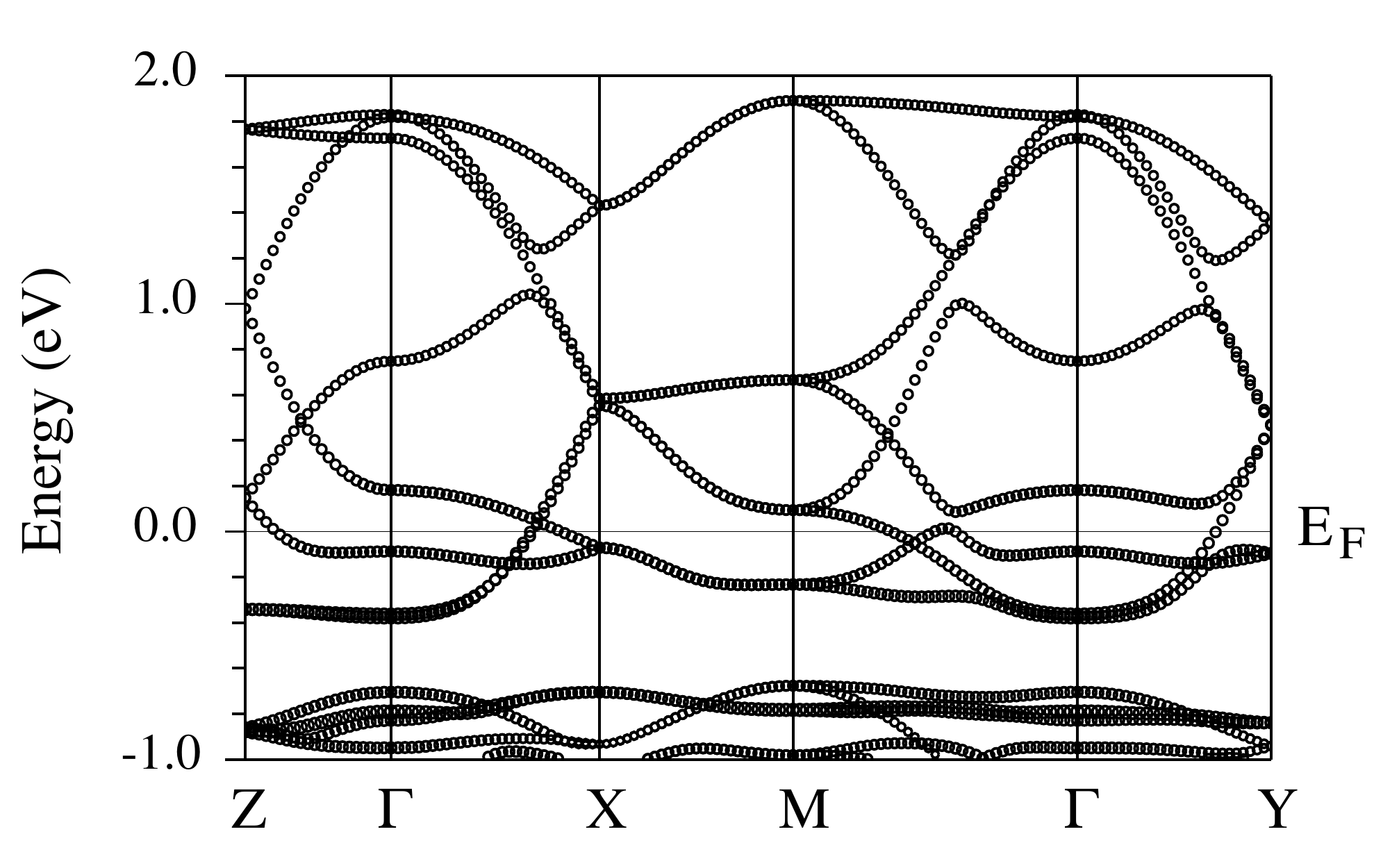}\\
\includegraphics[width=\columnwidth]{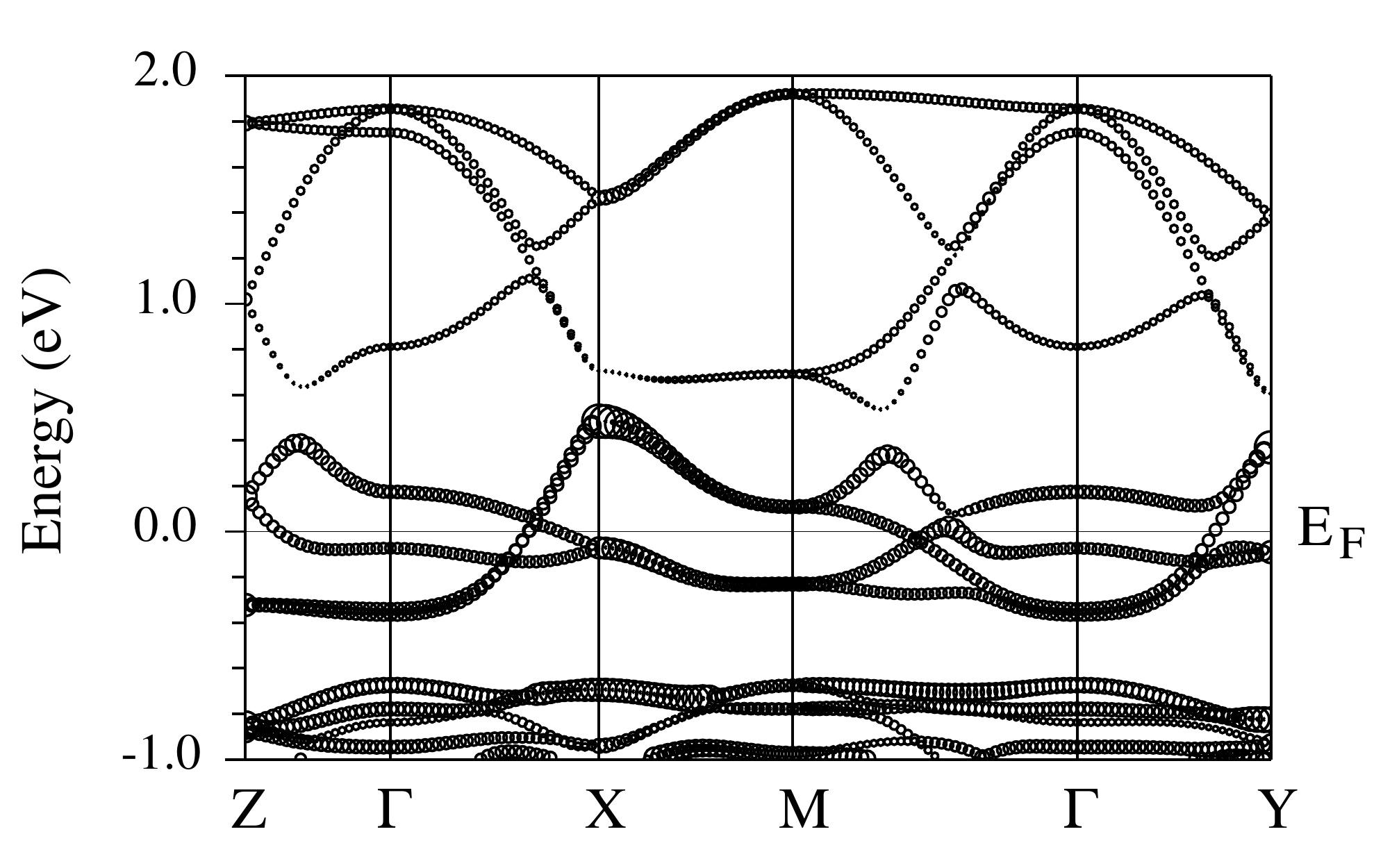}
\end{tabular}
  \caption{LDA band structures of \lunio in the high-temperature
    orthorhombic $Pbnm$ structure (top), and low-temperature
    monoclinic $P2_1/n$ structure (bottom). The Brillouin zone labels
    are $Z (0,0,0.5)$, $\Gamma (0,0,0)$, $X (0.5,0,0)$, $M
    (0.5,0.5,0)$, and $Y (0,0.5,0)$. A ``fat-band'' representation is
    used to display the Ni-LB site character. Larger dots denote
    larger Ni-LB character and smaller dots correspond to larger
    Ni-SB character.}
  \label{fig:bands}
\end{figure}

\begin{figure}
\begin{tabular}{c}
\includegraphics[width=\columnwidth]{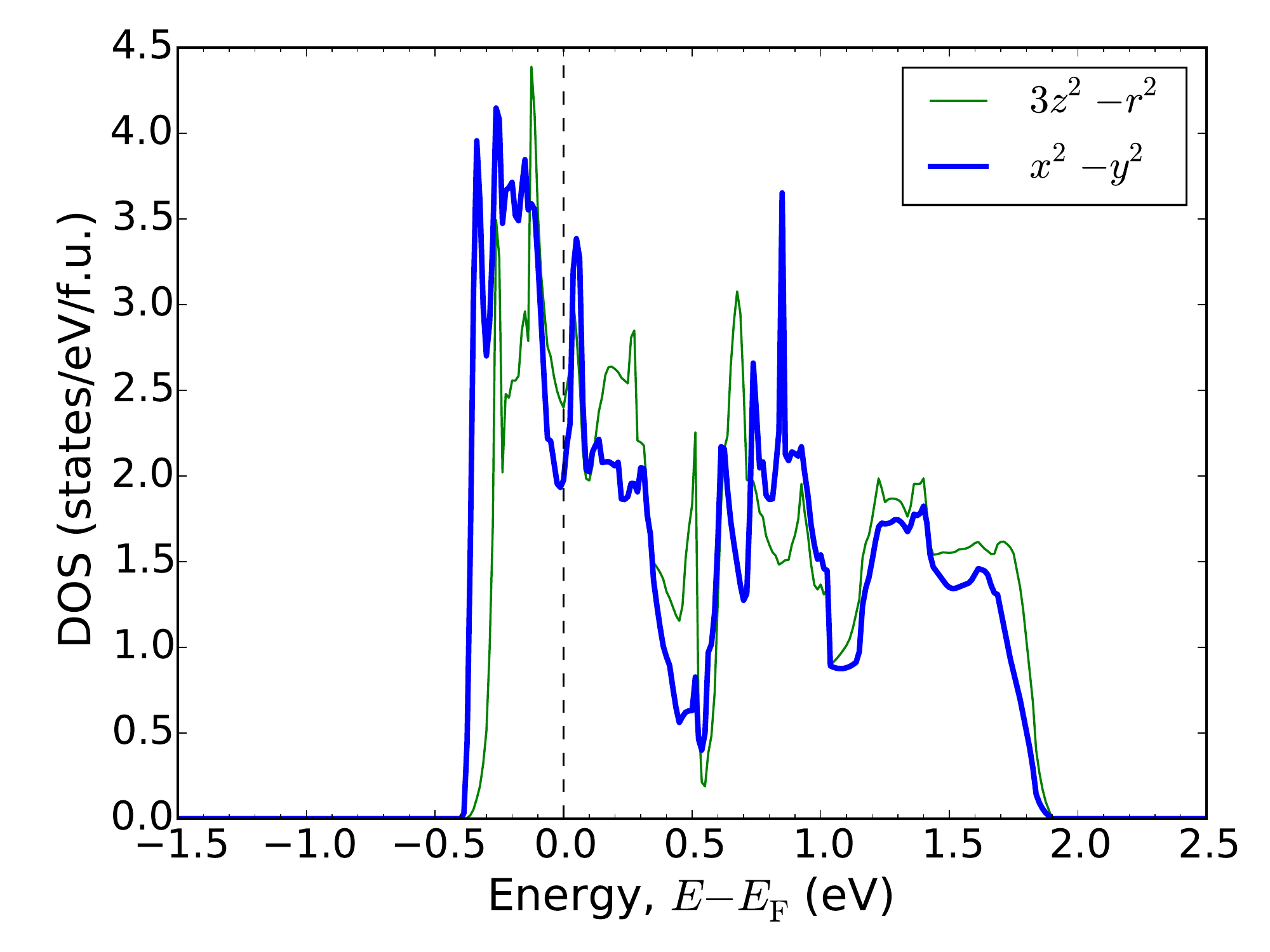}\\
\includegraphics[width=\columnwidth]{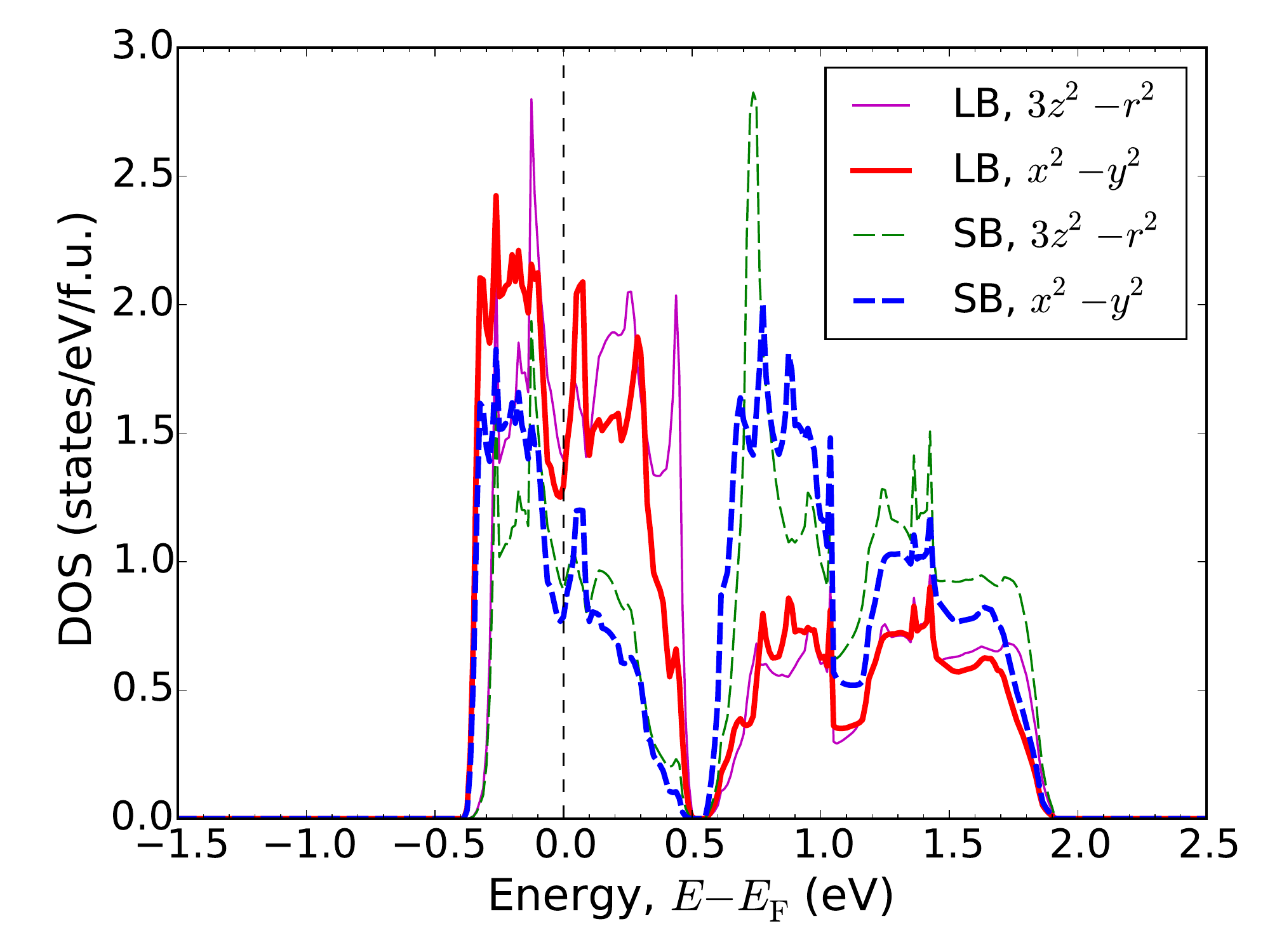}
\end{tabular}
  \caption{(Color online) Orbitally resolved LDA density of states of
    the $e_{g}$ bands for the orthorhombic (top) and monoclinic
    (bottom) structures.}
  \label{fig:dos_LDA}
\end{figure}

To determine the appropriate low-energy description for LuNiO$_3$, 
we first study its electronic structure using density-functional theory 
in the local density approximation (DFT-LDA). 
%

The high-temperature (metallic) phase of all rare-earth nickelates
that undergo an insulator-to-metal transition occurs in the GdFeO$_3$
type orthorhombic structure with the space group $Pbnm$. There are
four formula units per unit cell, corresponding to a $(\sqrt{2} \times
\sqrt{2} \times 2)$ enlargement with respect to the undistorted cubic
structure. The orthorhombic structure derives from the undistorted
structure following octahedral rotations 
of $a^-a^-c^+$ type in Glazer's notation~\cite{Glazer1972,Glazer1975}.
In the low-temperature insulating phase, there is a symmetry breaking
of the Ni sites into two sublattices such that the volume of the
NiO$_6$ octahedra in one sublattice increases, which simultaneously
results in the decrease of the volume of the corner-shared NiO$_6$
octahedra in the other sublattice.  The resulting structure is
monoclinic with the space group $P2_1/n$.
The volume of the long-bond (LB) octahedra is $10.69$~\AA$^3$, and that of the 
short-bond (SB) ones is $9.39$~\AA$^3$, a difference of $13$\%. The average 
Ni-O bond-length differs by $4.37$\%.

The calculated DFT-LDA bandstructure is displayed in
Fig.~\ref{fig:bands} for both the high-temperature orthorhombic
structure (top panel) and the low-temperature monoclinic structure
(bottom panel).
The corresponding density of states (DOS) of the $e_{g}$ bands are displayed in
Fig.~\ref{fig:dos_LDA}.
The details of the calculations are provided in
Appendix~\ref{sec:appendix}.
In both structures, the bands around the Fermi level derive from the
$e_g$ states, which are built out of strongly hybridized Ni $3d$ and O
$2p$ orbitals, reflecting the strong covalency in this material.
%
%
The $e_g$ manifold extends from $-0.4$ to $+1.9$~eV relative to the
Fermi level, corresponding to an overall bandwidth $\Wlarge\simeq
2.3$~eV, and is well separated from other bands (such as $t_{2g}$).
There are eight spin-degenerate bands within this manifold as there
are four Ni per primitive unit cell and two $\eg$-orbitals per Ni.

In the orthorhombic structure, the bands in the $e_g$ manifold cross
at isolated points within the energy window between $+0.4$ and
$+0.6$~eV.
Correspondingly, the DOS in Fig.~\ref{fig:dos_LDA} (top panel) displays a dip in 
this energy range.
The degeneracies at these crossing points are lifted in the monoclinic
structure, and the $e_g$ manifold is split into two distinct sub-manifolds, 
each comprising four bands. The lower partially occupied sub-manifold has 
a bandwidth $\Wsmall\simeq 1.0$~eV, significantly reduced as compared to $\Wlarge$. 
The two sub-manifolds are separated by a characteristic energy scale 
$\delsite\simeq 0.25$~eV. 

The physics underlying the splitting of the $\eg$ manifold into two subsets as 
one goes from the \ortho to the \mono structure is quite easily explained. 
There is smaller Ni $d$-O $p$ covalency in the LB octahedra because of
the larger interatomic distance, which pushes the $\eg$ states
corresponding to the LB octahedra to lower energies.  The opposite
applies to the SB octahedra.  Hence, the lower submanifold has a
stronger Ni-LB character, as well as a smaller bandwidth, while the
upper submanifold has a stronger Ni-SB character and a larger
bandwidth.  This is illustrated by the `fat-band' representation in
Fig.~\ref{fig:bands} used to display the Ni-LB site-character.
The splitting between the two submanifolds can be related to the difference of on-site energies 
(averaged over the two orbitals) between the LB and SB nickel sites.
We obtain (in eV): 
$\epsilon(\mathrm{Ni}_{LB1}) = 0.44, \epsilon(\mathrm{Ni}_{LB2}) = 0.52$,
and 
$\epsilon(\mathrm{Ni}_{SB1}) = 0.67, \epsilon(\mathrm{Ni}_{SB2}) = 0.79$, 
yielding 
$\delsite =  (0.79+ 0.67)/2 - (0.44 + 0.52)/2 = 0.25$~eV, in good agreement 
with the observed band-splitting. 
Because $\delsite$ is a rather small energy scale (in comparison, for
example, with $\Wsmall$), the two bands actually have a rather mixed
character of Ni-LB and Ni-SB, although at degeneracy points the lower
(upper) manifold is entirely LB (SB), as expected.  In the limit of
large $\delsite$ however, the lower manifold would be fully LB and the
upper one fully SB, with a smooth connection between the small and
large $\delsite$ limits.

Correspondingly, the charges (electronic occupancies) of the two types
of nickel sites are different. At the LDA level, we find $1.16$
electrons for the LB sites and $0.84$ electrons for the SB sites.  One
should be careful, however, in interpreting these numbers in
connection with charge disproportionation between the two types of
sites. We emphasize that these are the occupancies referring to the
strongly hybridized, extended, low-energy $\eg$ states.  The
occupancies of these states {\it cannot be} interpreted as the valence
state of the localized atomic-like nickel orbitals.

We finally observe that the energy difference between the $x^2-y^2$
and $3z^2-r^2$ orbitals on a given site is a smaller energy scale than
$\delsite$ and, of course, the bandwidth, emphasizing that the
Jahn-Teller effect plays little role.

\subsection{Low energy Hamiltonian}
\label{sec:ham}


In order to construct a low-energy Hamiltonian for the $\eg$ states,
we need a set of site-centered localized wave-functions describing
these states.
To this aim, we have built maximally localized Wannier functions using the 
procedure of Refs.~\cite{Marzari1997,Wannier90,Marzari2012_review}.
We used an energy window of $[-0.4, +2.0]$~eV that encloses the eight bands of the $e_g$ manifold 
and obtain two Wannier orbitals per each Ni site corresponding to the two $e_g$-like orbitals 
\footnote{
The spatial spreads which are minimized in the maximally localized Wannier came out to be  
Ni$_{LB1}$: 4.32, Ni$_{LB2}$: 4.42,
Ni$_{SB1}$: 4.17, Ni$_{SB2}$: 4.17 (in \AA$^2$).}.

We consider the following low-energy Hamiltonian, 
which contains a kinetic energy (band-structure) term $H_b$ and an interaction term $\Hint$: 
\begin{equation}
H\,=\,H_b\,+\sum_i\,H_{\mathrm{int}}(i) .
\label{eq:ham}
\end{equation}
The kinetic energy term reads: 
\begin{equation}
H_b\,=\,\sum_{\vk\sigma\nu}
\varepsilon_{\nu}(\vk)\,e^\dagger_{\vk\sigma\nu}e_{\vk\sigma\nu} .
\label{eq:ham_bands}
\end{equation}
In this expression, $\varepsilon_{\nu}(\vk)$ is the band dispersion
calculated above for either the orthorhombic or the monoclinic
structure. The eight different bands are labeled by the index $\nu$,
and the sum over pseudomomenta $\vk$ runs over the Brillouin zone of
each structure.  The operator $e^\dagger_{\vk\sigma\nu}$ creates an
electron in one of the $\eg$ bands with spin $\sigma$.

In Eq.~(\ref{eq:ham}), the index $i$ refers to sites of the crystal
lattice (with $4$ sites per unit cell and two types of inequivalent
sites in the monoclinic structure corresponding to the LB and SB
octahedra). On each site, we make the simplest possible choice for
$H_{\mathrm{int}}$, namely a Kanamori Hamiltonian, appropriate to two
orbitals per site with local interactions only:
\begin{eqnarray}
H_{\mathrm{int}}\,=\,U\sum_m \hnmu\hnmd\,+\,(U-2J)\sum_{m\neq\mp} \hnmu\hnpmd\,+&\nonumber\\
+(U-3J) \sum_{m<\mp,\sigma} \hn_{m\sigma}\hn_{\mp\sigma} +\,\,\,& 
\label{eq:ham_kanamori} \\
+ J\, \sum_{m\neq\mp} e^+_{m\spinup}e^+_{m\spindown}\,e_{\mp\spindown}e_{\mp\spinup} 
-J\,\sum_{m\neq\mp} e^+_{m\spinup}e_{m\spindown}\,e^+_{\mp\spindown}e_{\mp\spinup}
\notag
\end{eqnarray}
Here, the orbitals (labeled by $m$) refer to the site-centered Wannier
functions constructed above.
%
In our calculations we omit the spin-flip and pair-hopping terms for
computational efficiency. Although they are important for multiplet
degeneracies and magnetic properties, the shift of the phase
boundaries caused by omitting them is relatively small
\cite{Pruschke2005}. Therefore, we expect that such an approximation
does not affect the qualitative conclusions of this article.

\section{Phase transitions and crossovers}
\label{sec:phases}

\subsection{Basic strategy}
\label{sec:strategy}

Here we investigate the phase diagram of the above low-energy model,
as a function of $U$ and $J$, for the two band structures corresponding
to the orthorombic and monoclinic phases.
%
Because $U$ and $J$ are effective low-energy coupling constants, their
actual values are not known \textit{a priori}.
Our basic strategy is to find out whether a range of coupling exists
in which the orthorombic structure is metallic while the monoclinic
structure is insulating. This range should not be too narrow so that
the effective low-energy description does not rely on excessive
fine-tuning of the coupling constants. Furthermore, it should
correspond to a range of couplings which is physically reasonable.

In this article, we shall not address the issue of estimating the
low-energy values of $U$ and $J$ from first-principles
methods. Progress has been made recently in the determination of
screened Coulomb interaction parameters using methods such as
constrained RPA \cite{Aryasetiawan2004} and combinations of GW and
DMFT \cite{Biermann2003}.  Further work is needed however to assess
the reliability of these methods in constructing low-energy models of
late transition metal oxides, \textit{i.e.}\ for projecting out the
ligand states despite their strong hybridization.  This is of
particular concern here, since ligand holes and negative charge-transfer
physics are likely to play an important role.  Furthermore,
no application of these methods to nickelates have yet appeared in the
literature, to our knowledge.

Finally, we emphasize that we shall limit ourselves to the
non-magnetic phases since our focus is on the metal to paramagnetic
insulator transition of nickelates with heavier rare-earths. The
interplay with magnetism and the simultaneous insulating and magnetic
transition of Pr and Nd compounds is left for future investigations.

\subsection{Phase diagram}
\label{subsec:phasediag}

\begin{figure} 
\includegraphics[width=\columnwidth]{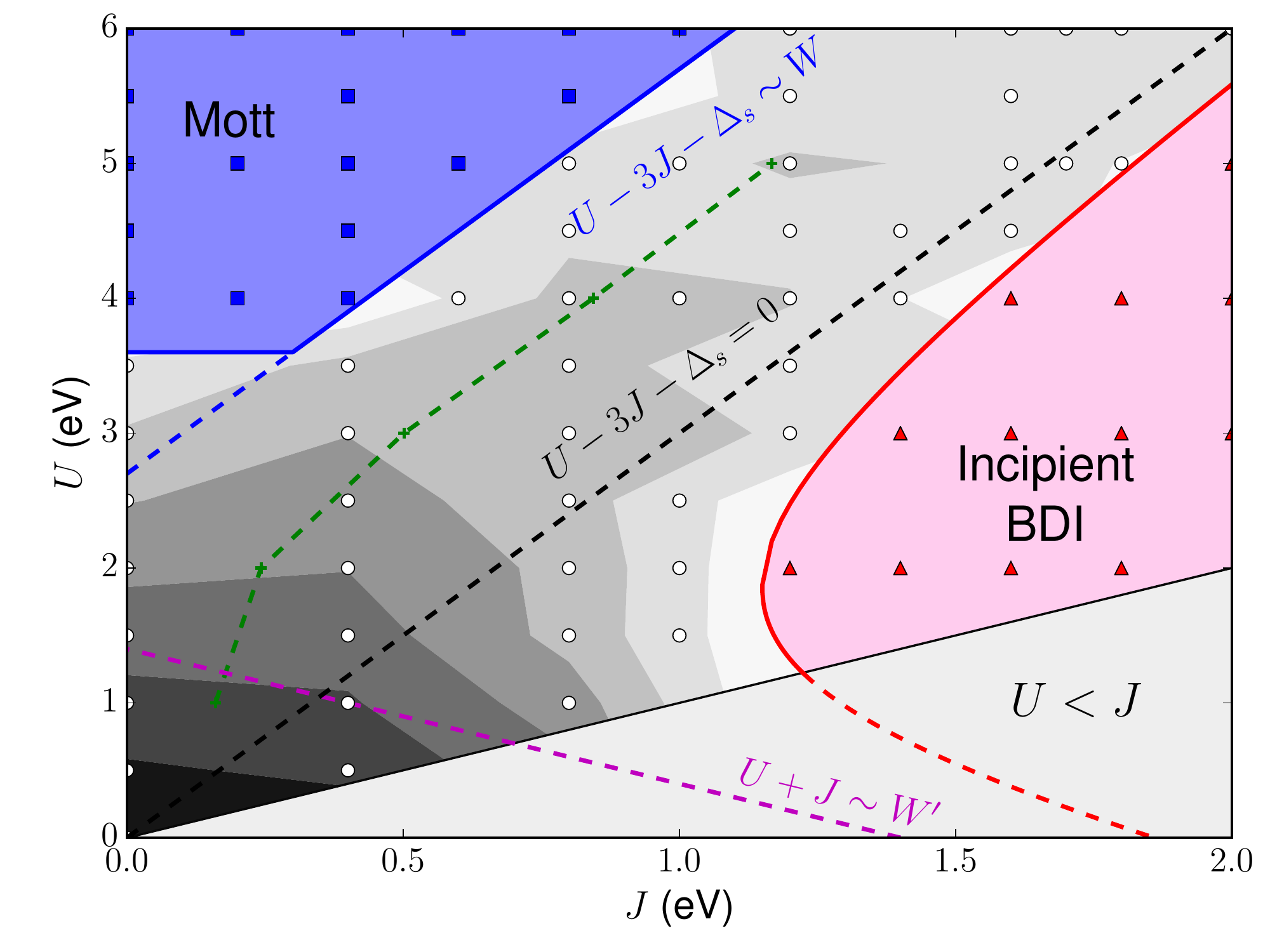}\\
\includegraphics[width=\columnwidth]{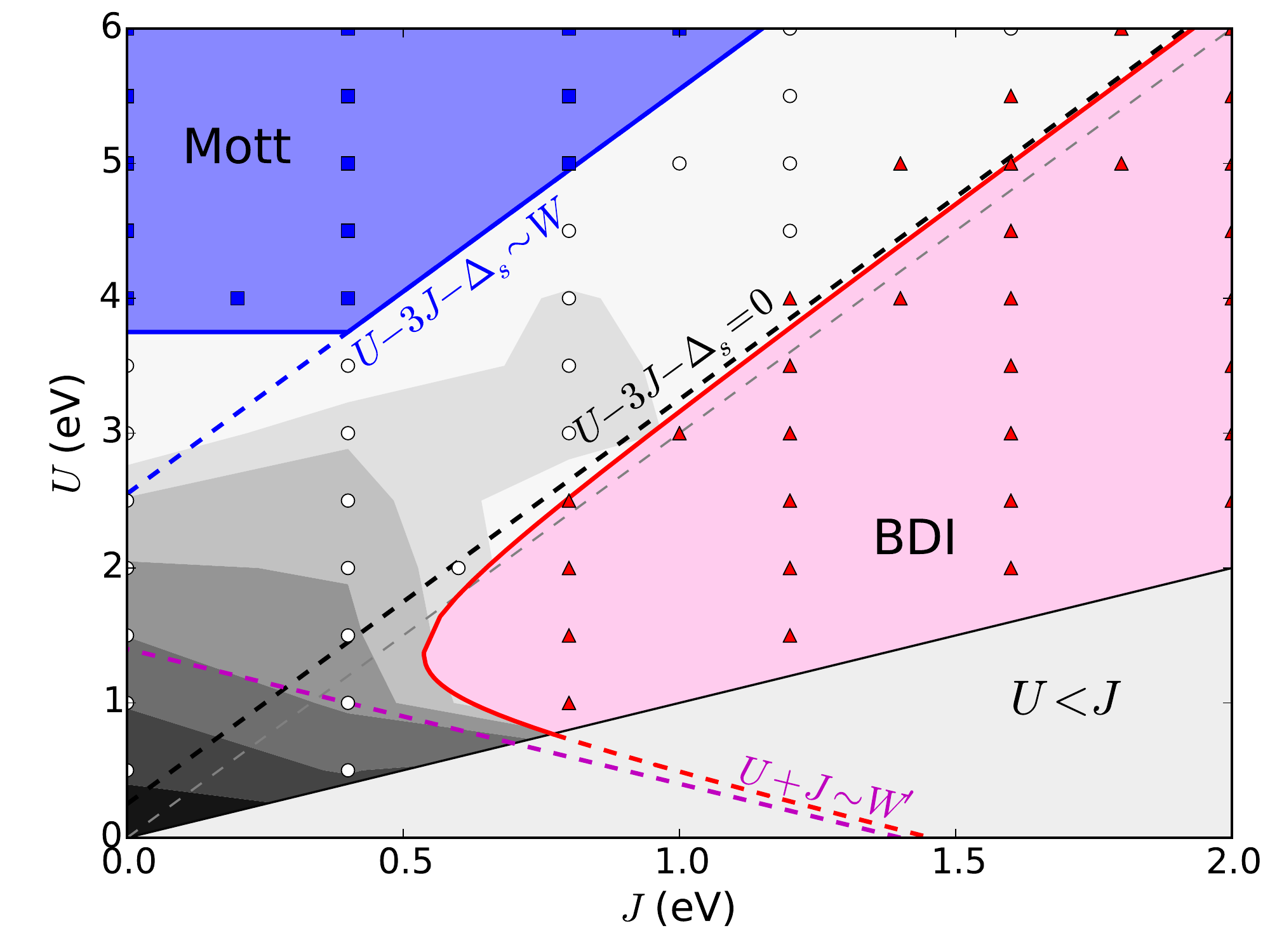}
  \caption{(Color online) Phase diagram of the low-energy model for
    \lunio as a function of Hund's rule coupling $J$ and on-site
    repulsion $U$.  Top: orthorhombic phase. Bottom: monoclinic phase.
    Straight lines separating various regions are designated according
    to estimates presented in the text. The green dashed curved line
    in the top panel show the locus of the maxima of the quasiparticle
    weight $Z$ as a function of $J$ as extracted from
    Fig.~\ref{fig:qpweight}.  The thin gray dashed line across the
    bottom panel represents a line $U - 3J = 0$ and is the same as the
    line $U - 3J - \delsite = 0$ for the orthorhobmic structure in the
    top panel (for which $\delsite = 0$).  The boundaries of the BDI
    phase are plotted according to Eq.~\eqref{eq:uc_j} with parameters
    $W$ and $W'$ fitted in order for the boundary to encompass the
    points obtained as insulating in DMFT calculations.  }
  \label{fig:phasediag}
\end{figure}

In Fig.~\ref{fig:phasediag}, we display the phase diagram of the
low-energy model for the orthorombic (top) and monoclinic (bottom)
structures as a function of $U$ and $J$. These results were obtained
using DFT+DMFT applied to the model given by
Eqs.~\eqref{eq:ham}-\eqref{eq:ham_kanamori} with the band structure
described in Section~\ref{sec:elstruc_bands}.  The details of the
calculations are provided in Appendix~\ref{sec:appendix}.

Both phase diagrams display two insulating phases separated by a
metallic phase.
In order to facilitate the understanding of the different regimes, it
is useful to consider the physically important crossover line
displayed as a black dashed straight line on the figure.
This line is defined by comparing the energies, in the atomic limit,
of the configuration in which two neighbouring nickel sites are each
occupied by a single electron ($\eg^1\eg^1$) versus the one in which
full disproportionation takes place ($\eg^2\eg^0$, with the two
electrons residing on the LB-nickel site). The energy of the former
configuration is zero, while the one of the latter is
$U-3J-\delsite$. This yields the following atomic-limit estimate:
\begin{equation}
U-3J\,=\,\delsite, 
\end{equation}
corresponding to the black dashed line displayed on both panels of
Fig.~\ref{fig:phasediag}.  For the upper panel, corresponding to the
orthorombic structure in which all sites are equivalent and
$\delsite=0$, this reduces to $U-3J=0$. Below this line the smaller
entry of the coupling constant matrix (\textit{i.e.}\ $U-3J$) becomes
negative, corresponding to the low-energy description of a small or
negative charge-transfer gap.  For the monoclinic phase of \lunio
(bottom panel), we have used our LDA estimate $\delsite\simeq 0.25$~eV
in drawing the crossover line.

The region of the phase diagram well above this crossover line is
quite conventional.  As $U$ is increased, a Mott transition is
encountered. For the orthorhombic structure, this is simply the Mott
transition of a quarter-filled two-orbital system corresponding to the
homogeneous filling of one electron per site. At $J=0$, we find this
transition to occur at $U^{(1/4)}_{c0}\simeq 4$~eV, about twice the
bandwidth $W_{>}$.
A non-zero value of the Hund's coupling increases this critical value,
in agreement with established
knowledge~\cite{georges_Hund_review_annrev_2013,vandermarel_sawatzky_prb_1988,
  demedici_MottHund,demedici_prl_2011}, roughly according to
$U_c=U^{(1/4)}_{c0}+3J$ (indicated as a straight blue line on the
figure).
For the monoclinic structure, the location of the Mott critical
boundary is not very significantly affected. This is expected (see Sec.~\ref{sec:discussion}) 
in view of the rather small value of $\delsite$: in the regime
$U-3J\gg\delsite$, the difference in the on-site energy due to the
bondlength disproportionation is a small effect and the kinetic energy
is still set by the full bandwidth $W_{>}$.
In particular, for a typical physical value of $J\simeq 0.8$~eV likely
to be appropriate for nickelates, we observe that the Mott transitions
of the two structures occur at values of $U$'s which are quite close
to each other. 
As a result, explaining that the \ortho
structure is metallic while the \mono one is insulating would require
fine-tuning of the interaction parameters.
Furthermore, in this regime the large $U$ suppresses the charge
disproportionation that is already present at the DFT level, yielding
a Mott insulating state that is qualitatively like the Mott insulating
solution for the orthorhombic phase.
Hence, we conclude that the regime $U-3J\gg\delsite$ (above the dashed
black crossover line) is not the relevant one for a proper low-energy
description of nickelates.
This theoretical consideration provides clear support to the fact that
the metal-insulator transition of nickelates cannot be viewed as the
homogeneous Mott transition of a quarter-filled band, as mentioned in
the introduction.

Let us now turn our attention to the regime $U-3J\lesssim \delsite$ on
the lower side of the crossover line.  There we see that, for large
enough $J$, an insulating phase is obtained, which is characterized by
a significant disproportionation of the $\eg$ occupancy between two
neighboring nickel sites. For this reason, this phase is labeled `bond
disproportionated insulator' (BDI) on the phase diagram of
Fig.~\ref{fig:phasediag} (one may also refer to it as a
`site-selective Mott insulator' as in \cite{Park2012}, or as a
`hybridization-wave insulator').

For the monoclinic structure, where the symmetry between the two
nickel sublattices is explicitly broken, this BDI phase has a much
larger extension. It basically covers the whole triangular-shaped area
contained between an upper boundary, which happens to be very close to
the crossover line $U-3J=\delsite$, and a lower boundary defined by
the straight line $U+J\simeq W_{<}$ (negative-slope dashed magenta
line in Fig.~\ref{fig:phasediag}).  The latter expression is again
easily rationalized by considering the atomic limit, this time for a
{\it half-filled} two-orbital shell occupied by two
electrons~\cite{georges_Hund_review_annrev_2013}.  Indeed, in this
case, the effective atomic gap is
$U_{\mathrm{eff}}=U+J$~\footnote{Denoting by $E(N)$ the lowest energy
  of the configuration with $N$ electrons in the atomic limit, one
  has: $U_{\mathrm{eff}}\equiv
  [E(3)-E(2)]-[E(2)-E(1)]=U+U-2J-[U-3J]=U+J$.}, and the relevant
bandwidth is $W_{<}$ corresponding to the lower band manifold.

For the orthorhombic structure, the incipient BDI phase corresponds to
a {\it spontaneous} breaking of the symmetry, in which two
inequivalent nickel sublattices occur (see Fig.~\ref{fig:ndisp}
below).  In this phase a hybridization wave (or a modulation of
covalency) develops on Ni-O bonds, which in reality would immediately
lead to a corresponding modulation of bond lengths.
It is interesting indeed that a homogeneous solution with all
equivalent nickel sites becomes unstable when $U-3J$ is too negative
(and $U+J$ exceeds a critical value).  This points to the extreme
sensitivity of the system to disproportionation in the `small or
negative charge-transfer' regime.  It also implies that lattice
degrees of freedom are coupled to the electrons in an essential way in
this regime.

An important observation, comparing the top and bottom panels of
Fig.~\ref{fig:phasediag}, is that the area covered by the BDI phase in
the \ortho case is significantly smaller than in the \mono case.
Although $\delsite$ is small, it shifts the BDI boundary to the left 
by an appreciable amount.  Hence, there is a rather extended region of
coupling constants $(U,J)$ in which the orthorhombic structure is
metallic, while the monoclinic one is an insulator (BDI).  We propose
that this is the physical region appropriate for a low-energy
description of nickelates.
This does not require 
fine-tuning of the low-energy coupling constants. 
In Sec.~\ref{sec:exp}, we will describe in more detail the behavior of
several physical quantities obtained for the values $J=0.8$~eV and
$U=1.0$ or $U=2.0$~eV, which lie within this region. These values were
chosen for illustrative purposes and correspond to a rather standard
value of $J$ for nickelates, leaving a more accurate determination of
the appropriate low-energy parameters for future investigations.

\begin{figure} 
\includegraphics[width=\columnwidth]{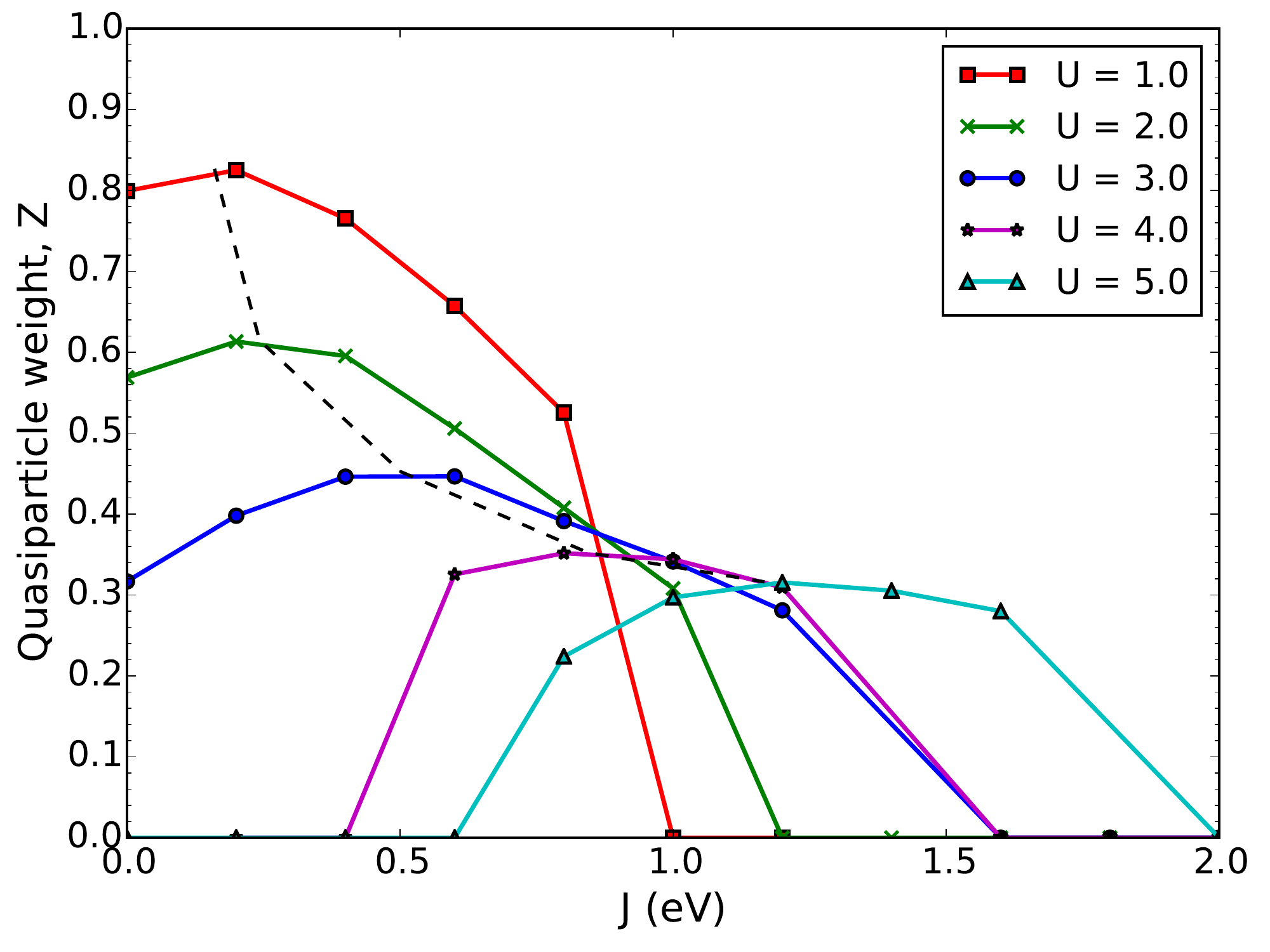}
  \caption{(Color online) Quasiparticle weight $Z$ (averaged over
    sites and orbitals) as a function of $J$ for different values of
    $U$ in the \ortho structure. Dashed line runs through the maxima
    of $Z(J)$. The locus of the maxima is also indicated as a green
    dashed line in the top panel of Fig.~\ref{fig:phasediag}. }
  \label{fig:qpweight}
\end{figure}

Finally, we briefly discuss the metallic phase separating the two
insulating regions (Fig.~\ref{fig:phasediag}, top panel).  In
Fig.~\ref{fig:qpweight}, we display the quasiparticle weight 
$Z$~\footnote{The quasiparticle 
weight $Z$ is calculated from the real-frequency self-energy  
as $Z = [1 -\partial\textrm{Re}\Sigma(\omega)/\partial\omega|_{\omega\to0}]^{-1}$.}  
throughout this metallic phase, as a function of $J$, for several
values of $U$.  Correspondingly, an intensity map of $Z$ as a function
of $U$ and $J$ is displayed in Fig.~\ref{fig:phasediag} using
different levels of grey-shading.
The quasiparticle weight $Z$ decreases as both the Mott phase and the
BDI phase are approached (Fig.~\ref{fig:phasediag}).  As a result, it
displays a non-monotonous behaviour as a function of the Hund's rule
coupling $J$ (Fig.~\ref{fig:qpweight}).  Well to the left of the
crossover line $U-3J=0$, $Z$ increases as $J$ is increased. This is
indeed the expected behavour for a quarter-filled correlated
metal~\cite{georges_Hund_review_annrev_2013}. In contrast, as the
disproportionation line $U-3J=0$ is approached, $Z$ passes through a
maximum and then decreases as $J$ is increased, in line with the
behaviour of a half-filled correlated
metal~\cite{georges_Hund_review_annrev_2013}.  The location where $Z$
is maximum in the $(U,J)$ plane is indicated as a dashed curve on
Fig.~\ref{fig:phasediag}, which happens to lie well to the left of the
disproportionation crossover line.  This behaviour also implies,
interestingly, that at larger $J$ the magnitude of $Z$ increases as
$U$ is increased.
We note in passing that the parameter set $U=2,J=0.8$~eV for which
more detailed calculations will be presented later in this article
corresponds to a quasiparticle weight (inverse mass enhancement)
$Z\simeq 0.35$ (Fig.~\ref{fig:qpweight}), a reasonable value for
metallic nickelates \cite{Rajeev1991,Sreedhar1992}.

In the larger $J$ regime, the BDI phase is surrounded by a metallic
phase both on the small $U$ side and larger $U$ side.  Hence, as $U$
is increased from weak-coupling in this regime, one encounters
successively: a metallic phase at small $U$, the BDI insulator, a
metallic phase again, and finally the Mott insulating phase at large
$U$.
The physical nature of the metallic phase in the `small or negative
charge-transfer regime' $U-3J\lesssim 0$ definitely deserves further
investigation, using e.g.\ DMFT techniques. This is relevant, in
particular, for a proper low-energy description of metallic LaNiO$_3$.

\section{Discussion and Qualitative Insights}
\label{sec:discussion}

\begin{figure*} 
\includegraphics[width=0.48\linewidth]{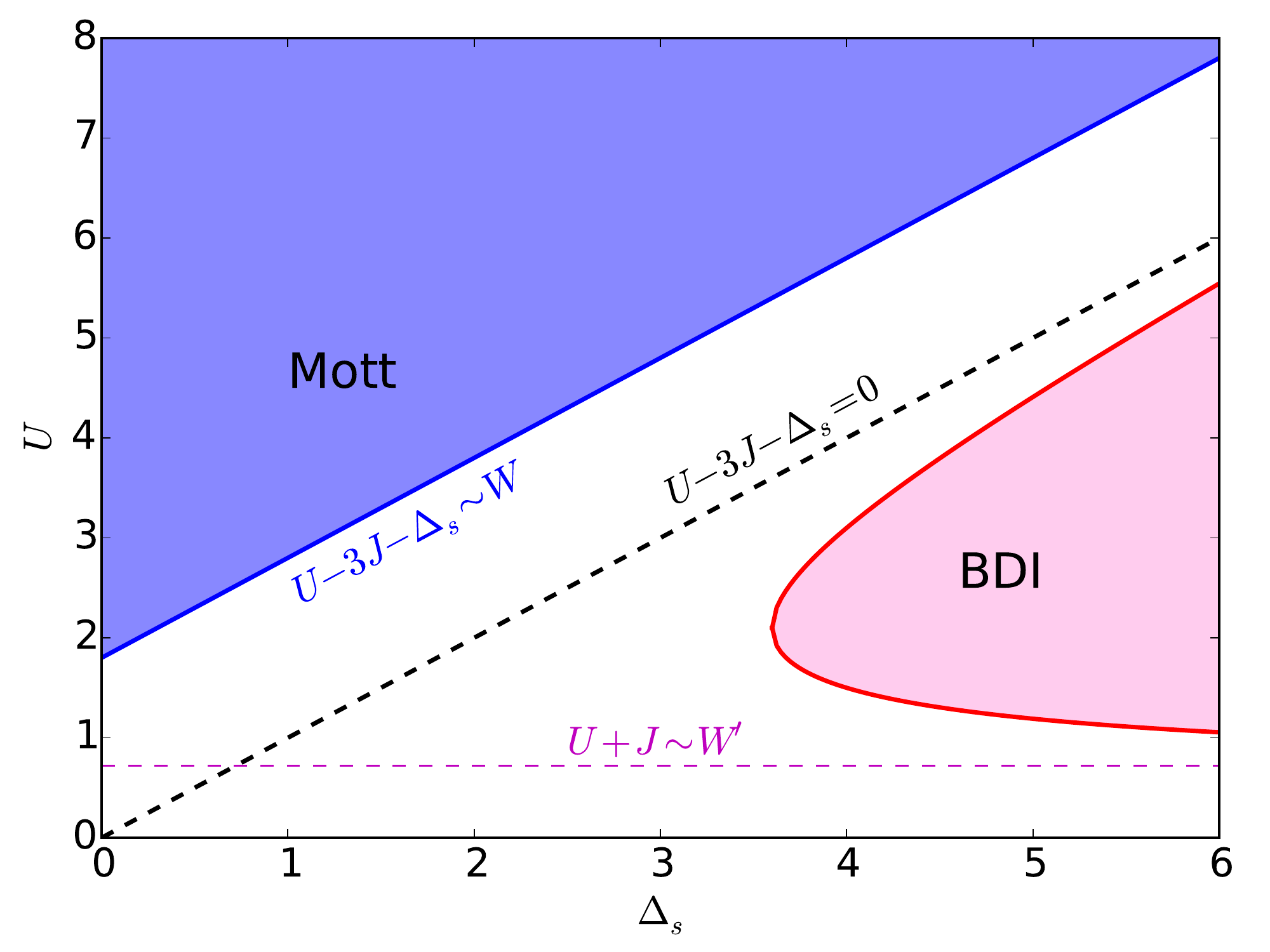}
\includegraphics[width=0.48\linewidth]{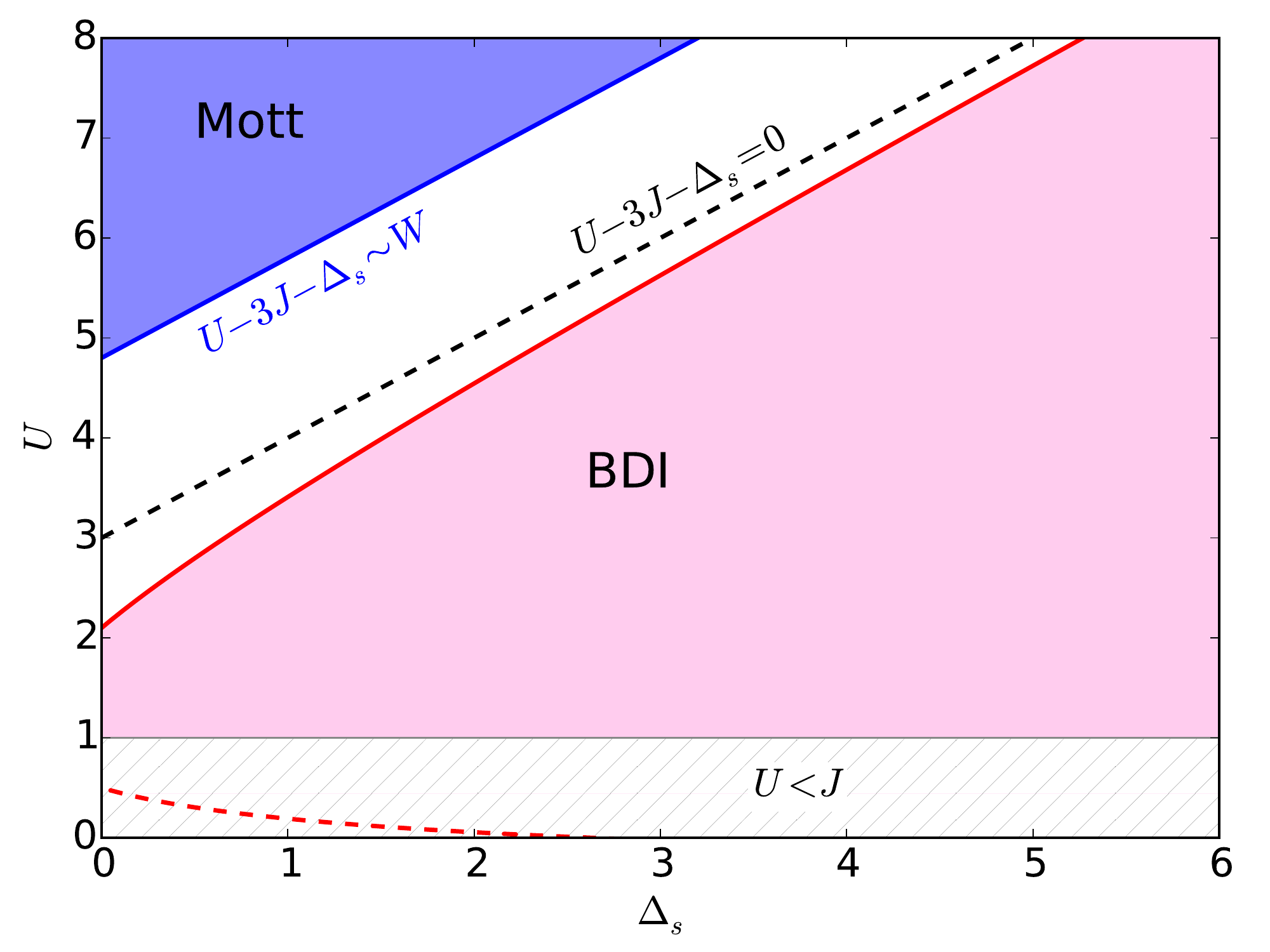}\\
\includegraphics[width=0.48\linewidth]{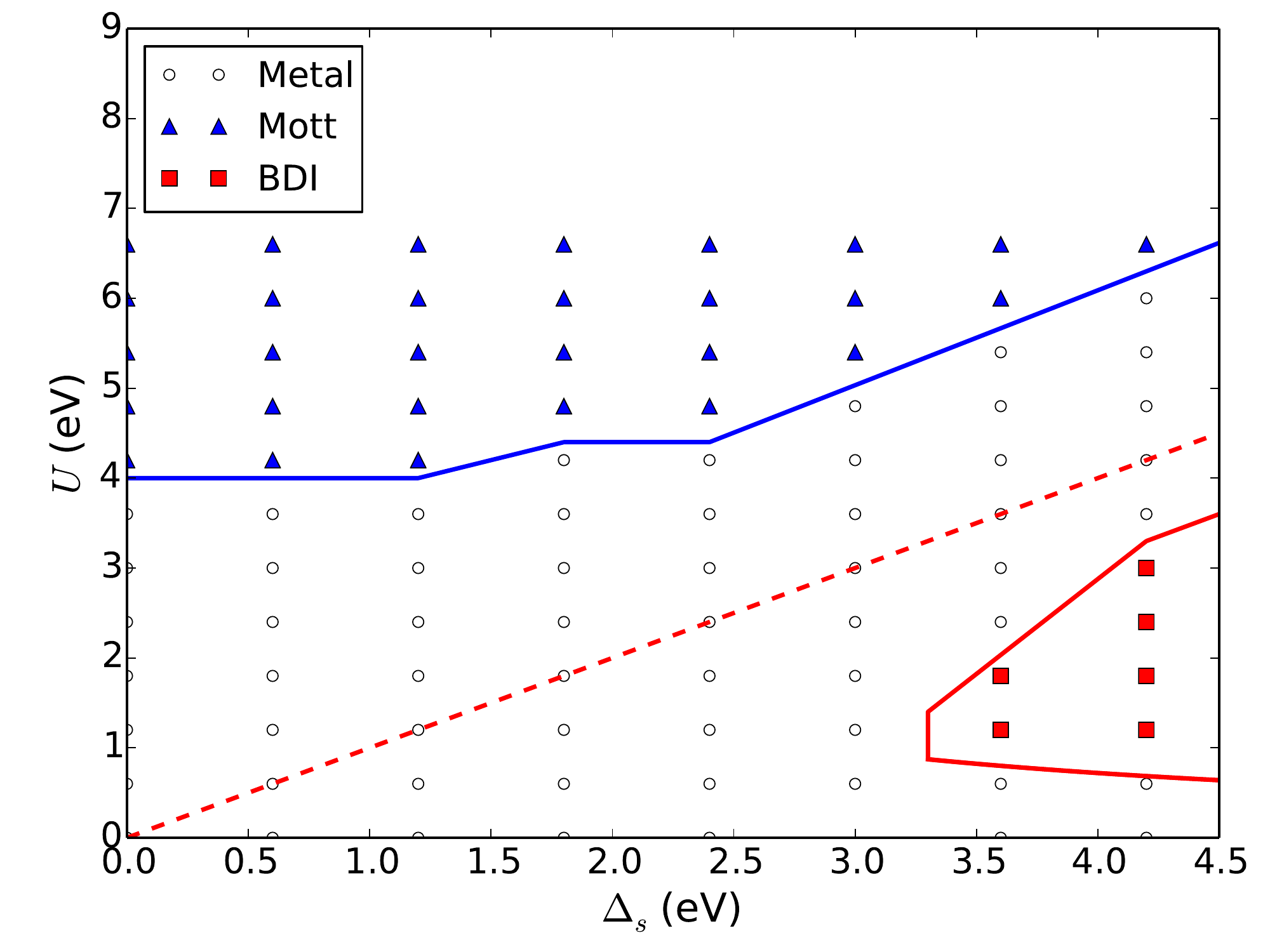}
\includegraphics[width=0.48\linewidth]{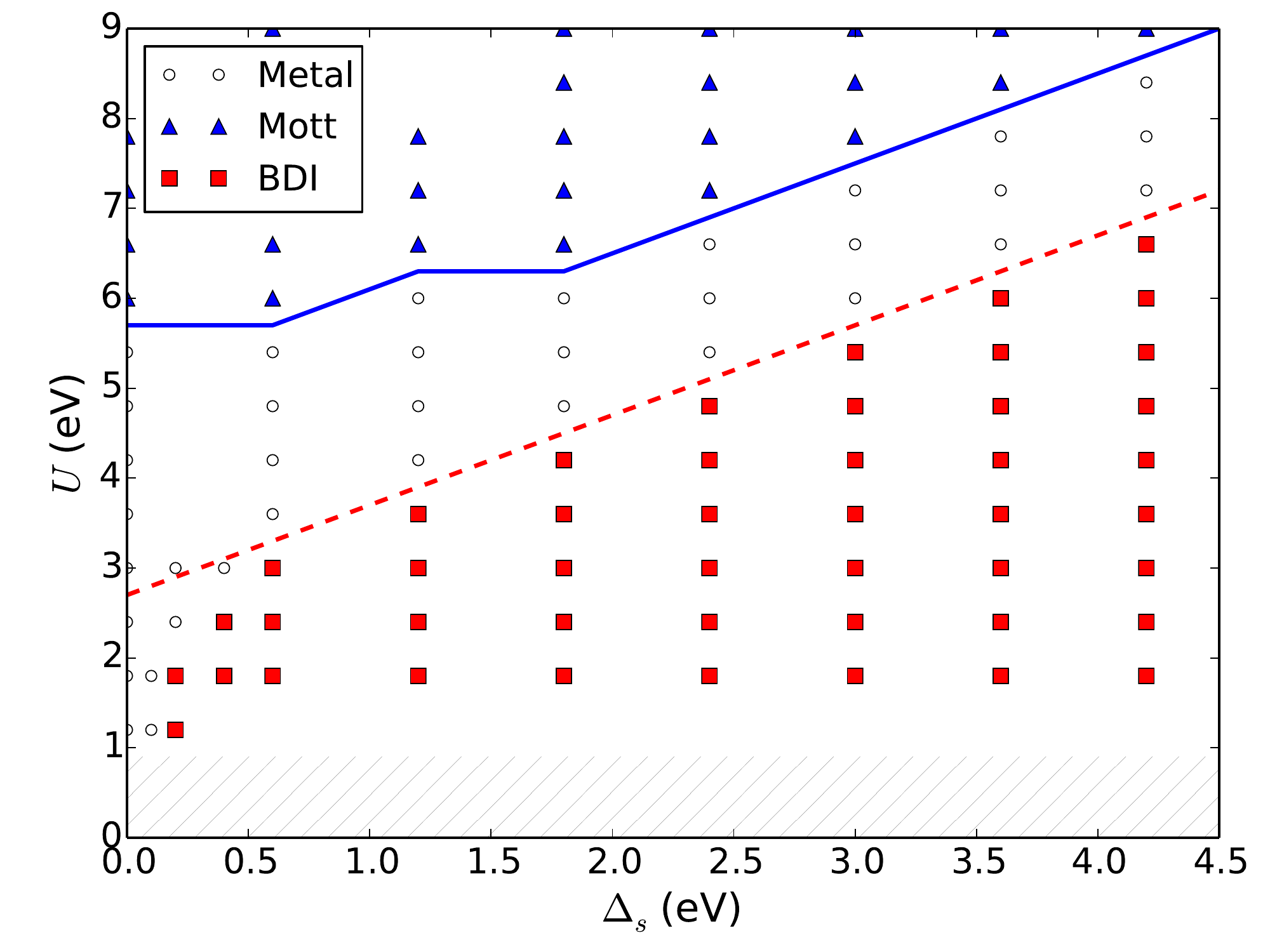}
  \caption{(Color online) Phase diagram of the two-orbital
    two-sublattice model. Top: Schematic phase diagrams based on
    qualitative considerations (see Appendix~\ref{appendix:model}) for
    $J < J_{c}$ (left) and $J > J_{c}$ (right).  Bottom: Actual phase
    diagrams obtained from DMFT calculations on a Bethe lattice, for
    $J = 0.0$ (left), and $J = 0.9$ eV (right). The dashed red line
    corresponds to $U - 3J - \delsite = 0$. The phase boundaries are
    indicative, with actual data being represented by the markers.}
  \label{fig:model}
\end{figure*}

The above analysis of the two structures of \lunio reveal that even though $\delsite$ 
is a small energy scale, it has a large effect on the phase diagram: the system 
is extremely sensitive to disproportionation in the regime when $U - 3J \lesssim 0$.

To better understand the origin of this sensitivity, we consider a
simple model that captures the main features of the real system.
The model involves two sublattices with on-site energies split by $\delsite$. 
Each site carries two orbitals, and the Hamiltonian reads:
\begin{align}
H = & -t \sum_{m=1}^{2}\sum_{\sigma=\uparrow,\downarrow}\sum_{\langle ij\rangle}\, 
(d^{\dagger}_{m\sigma i} d_{m\sigma j} + h. c.) + H_{\mathrm{int}}\notag\\
 & - \frac{\delsite}{2} \sum_{m\sigma, i\in A} d^{\dagger}_{m\sigma i} d_{m\sigma i} +
 \frac{\delsite}{2}\sum_{m\sigma, j\in B} d^{\dagger}_{m\sigma j} d_{m\sigma j} ,
\label{eq:ham_model}
\end{align}
where the interaction Hamiltonian is the same as before as given by
Eq.~\eqref{eq:ham_kanamori}.  The difference in on-site energies
induced by bond disproportionation in nickelates is described here by
an explicitly introduced parameter $\delsite$. This allows us to study
how physical properties change as $\delsite$ is varied.  In reference
to nickelates, we will refer to sites $A$ and $B$ as LB and SB sites,
respectively.  We denote by $W$ the non-interacting bandwidth when
$\delsite=0$ ($W=4t$ for a semi-circular density of states).

Before proceeding to the results of DMFT calculations, let us perform
a qualitative analysis of this model.
Consider first the case of $J = 0$. A cartoon of the phase diagram is
sketched in the top left panel of Fig.~\ref{fig:model}, with the phase
boundaries being derived from qualitative estimates presented in more
details in Appendix~\ref{appendix:model}.
In the limit of large $\delsite$ ($\gtrsim U$) the model can be viewed
as a half-filled lattice of doubly occupied LB-sites connected via
indirect hopping $\teff$ mediated by empty SB-sites.
A key insight is that this effective hopping depends on the value of
$U$.  A simple estimate from second order perturbation theory around
the atomic limit (appendix \ref{appendix:model}) yields:
\begin{equation}
\teff\,\sim\,\frac{t^2}{\delsite-U}
\end{equation}
Increasing the Coulomb repulsion $U$ thus has two antagonistic
effects.  On the one hand, it induces a Mott transition in this
half-filled system. On the other hand, it decreases the energy
separation between the LB and SB sites, thus enhancing $\teff$ and
driving the system towards a metallic state. Such an ambivalent effect
of $U$ results in a reentrant behaviour of the metal-BDI-metal
transition apparent in Fig.~\ref{fig:model}: the metal to BDI critical
boundary has two branches, a lower one (metal $\rightarrow$ BDI)
with critical $U_c = \Ucs$ and an upper one (BDI $\rightarrow$ metal)
with critical $U_c = \Ucl$.
A simple estimate for these critical values can be obtained by writing that the critical 
boundary corresponds to $U$ of the order of the effective bandwidth: 
$U_c\simeq W^2/(\delsite-U_c)$. Hence, the upper branch of the critical boundary, for 
which $\delsite-U_c$ is small, is given by: 
$\Ucl\simeq\delsite-W^2/\delsite$, while the lower branch for which 
$\delsite\gg U_c$ is given by: $\Ucs\simeq W^2/\delsite$. 
For more detailed expressions, see Appendix~\ref{appendix:model}. 
An important point is that there is a critical value $\delc$ of the site disproportionation energy 
for which $\Ucs=\Ucl$, corresponding to the tip of the lobe of the BDI region on the upper 
left panel of Fig.~\ref{fig:model}. Equating the two expressions above, we see that 
$\delc$ is proportional to the bandwidth $W$ (when $J=0$). 
For $\delsite < \delc$, no transition to the BDI phase occurs and the system 
remains metallic 
(until eventually the Mott phase is reached). 
%

The regime $U \gtrsim \delsite$ has an entirely different physics, dominated by the 
Mott transition of a quarter-filled band.  The ground-state 
is now characterized by both LB and SB sites filled by one electron,
with the hopping parameter being simply the bare hopping $t$.
The lowest-energy excitation is determined by the energy of moving
an electron from a SB to a LB, which results in the effective interaction energy
$\Ueff = U - \delsite$ 
(while the energy cost of moving an electron from a LB to a SB site is $U + \delsite$).
The critical value of $\Ueff$ is then determined by the bandwidth $W$, which
leads to $U_{c} \simeq \delsite + W$ (boundary of the Mott phase in 
the top left panel of Fig.~\ref{fig:model}). 

%
Let us now consider a finite Hund's coupling $J>0$ which, as we shall see, has a major effect 
on the phase diagram (Fig.~\ref{fig:model}, top right panel). 
The crossover line between the quarter-filled and half-filled regimes (which is 
also the upper asymptotic boundary of the BDI phase at large $\delsite$) 
now reads $U - 3J - \delsite = 0$, as discussed in the previous section. 
Correspondingly, the Mott critical boundary in the quarter-filled region 
$U-3J >\delsite$ is simply shifted according to $U_{c}-3J-\delsite \simeq W$. 
More drastic changes occur in the half-filled regime $U-3J\lesssim\delsite$. 
Indeed the effective Coulomb energy in this regime is $U+J$, so that the metal to BDI 
phase boundary is now given by the criterion: 
$U+J \sim W^2/[\delsite-(U-3J)]=W^2/[\delsite+4J-(U+J)]$. 
This is the same expression as the one for $J=0$, but with crucial renormalizations of the 
parameters: $U\rightarrow U+J\,,\,\delsite\rightarrow \delsite+4J$. 
Hence, the Hund's rule coupling {\it effectively increases} the site disproportionation energy by $4J$. 
critical disproportionation is dramatically reduced down to
$\delc\simeq W-4 J$, as compared to the $J=0$ case.  As a result,
there is a critical value of the Hund's coupling, $J_{c} \simeq W/4$, such that
for $J>J_c$ the BDI instability takes place already at $\delsite=0$,
which means that in this regime the small or negative charge-transfer
metal is spontaneously unstable to site disproportionation.

Let us now turn to numerical simulations of
Hamiltonian~\eqref{eq:ham_model}.  We performed calculations within
DMFT using for simplicity a Bethe lattice (i.e. a semi-circular
density of states of width $W=4t$ when $\delsite=0$). In this case,
the DMFT self-consistency relation reads, for each sublattice:
\begin{eqnarray}
{\cal G}_{0A}^{-1} = i\omega_n+\mu+\frac{\delsite}{2}-t^2 G_B\\
{\cal G}_{0B}^{-1} = i\omega_n+\mu-\frac{\delsite}{2}-t^2 G_A,
\end{eqnarray}
where $\mathcal{G}_{0i}$ ($i=A,\,B$) are the Green's function of the
self-consistent bath for each site, and $G_{i}$ the corresponding
local (impurity) Green's function. The latter are obtained from
$\mathcal{G}_{0i}$ using the CT-QMC solver. The hopping parameter
$t=0.6$ eV is chosen to give the bandwidth value $W = 2.4$ eV, very close to
that of nickelates. The temperature is set to $T = t / 24 = 1/40$ eV$^{-1}$.

The resulting phase diagrams for $J = 0$ and $J = 0.9$ eV (a value
relevant for the nickelates) are presented in the bottom left and right
panels of Fig.~\ref{fig:model}, respectively.  Apart from some
details, the numerical results agree very well with the qualitative
analysis performed above, especially regarding the following features:
i) the quarter-filled (Mott) and half-filled (BDI) regimes are
separated by the crossover line $U - 3J - \delsite = 0$; ii) there is
a critical value $\delc$ of the on-site disproportionation energy
below which the system undergoes only a single, Mott-like, transition
and above which there are generally three transitions as a function of
$U$; iii) the value of $\delc$ is very sensitive to the Hund's
coupling $J$.  At some critical value of the Hund's coupling $J_{c}$
(corresponding to $\delc=0$), the phase diagram becomes qualitatively
different from that of the $J=0$ case.  In particular, the metallic
phase at $U < \Ucl$ becomes unstable with respect to the BDI phase for
all values of $\delsite$. Only when sufficiently large $U$ suppresses
the site disproportionation is the metallic phase stable, before
eventually turning into the Mott phase as $U$ is increased further.

In relation to the nickelates, the model calculations reveal that the
mechanism underlying the large sensitivity to $\delsite$ observed in
Fig.~\ref{fig:phasediag} is, in fact, tightly related to the effect of
the Hund's coupling on changing the position of the BDI phase
boundary.  We suggest that the nickelates are poised close to the
value $J\simeq J_c$.  In this regime, the system is highly sensitive
to disproportionation, even for small $\delsite$.  The estimate
$J_c\simeq W/4$ makes this a qualitatively reasonable range of
parameters.
This ties together two observations previously made in the literature:
the relevance of the small or negative charge-transfer
regime~\cite{Mizokawa2000} and the importance of the Hund's rule
coupling~\cite{Mazin2007}.
%

\section{Experimental implications: spectroscopy, local magnetic probes}
\label{sec:exp}

%

In this section, we discuss some physical implications of our
low-energy description in connection with experiments.
We do not aim at being quantitative in the present article. Indeed,
the appropriate values of the low-energy parameters $U$ and $J$ will
have to be determined ultimately by quantitative comparison to
experimental data, which we leave for future work. Rather, we focus
here on the main qualitative points.

\subsection{Spectroscopy: photoemission, optics}

\subsubsection{Single-particle spectral functions}

\begin{figure} 
\includegraphics[width=\columnwidth]{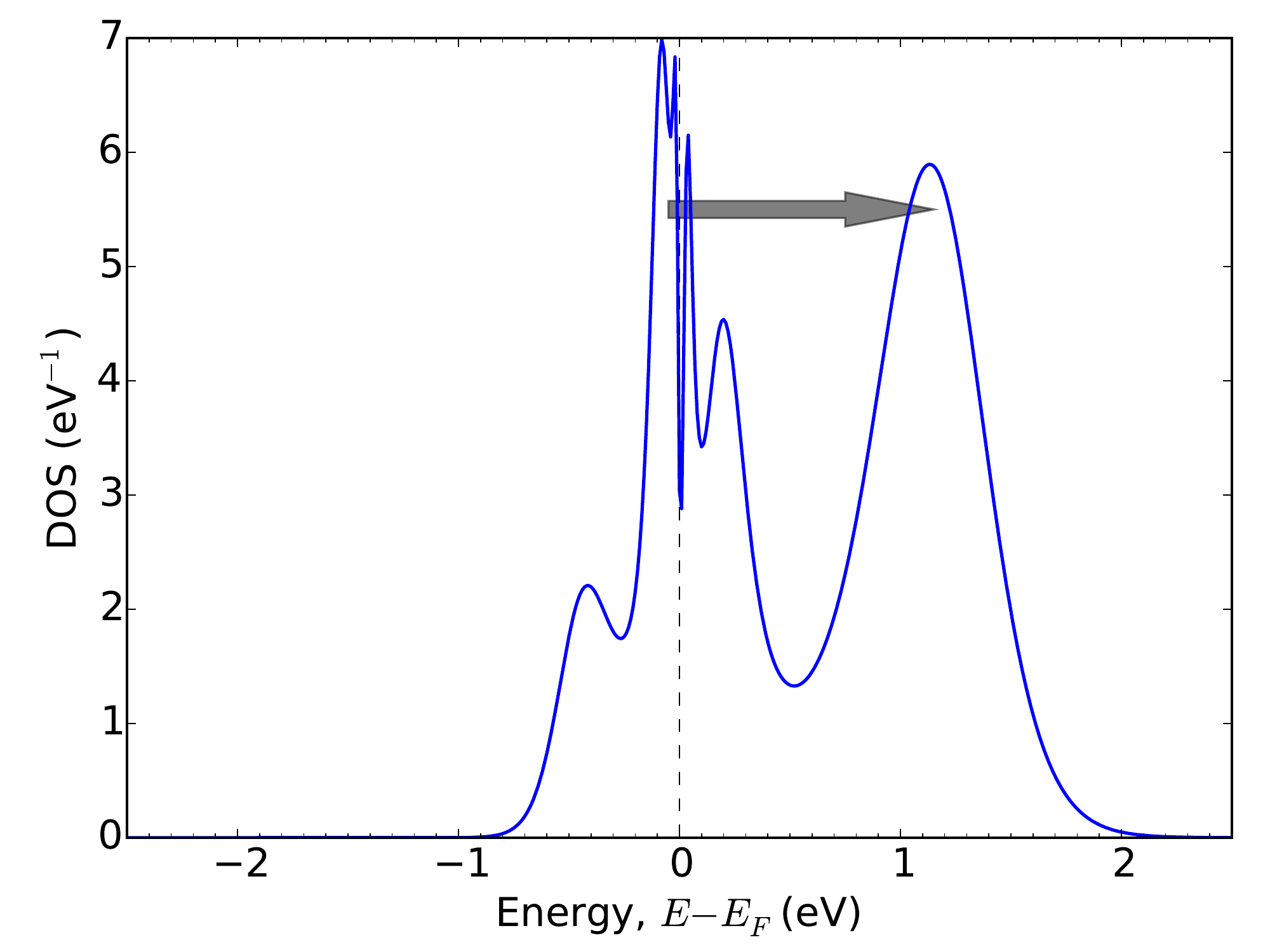}\\
\includegraphics[width=\columnwidth]{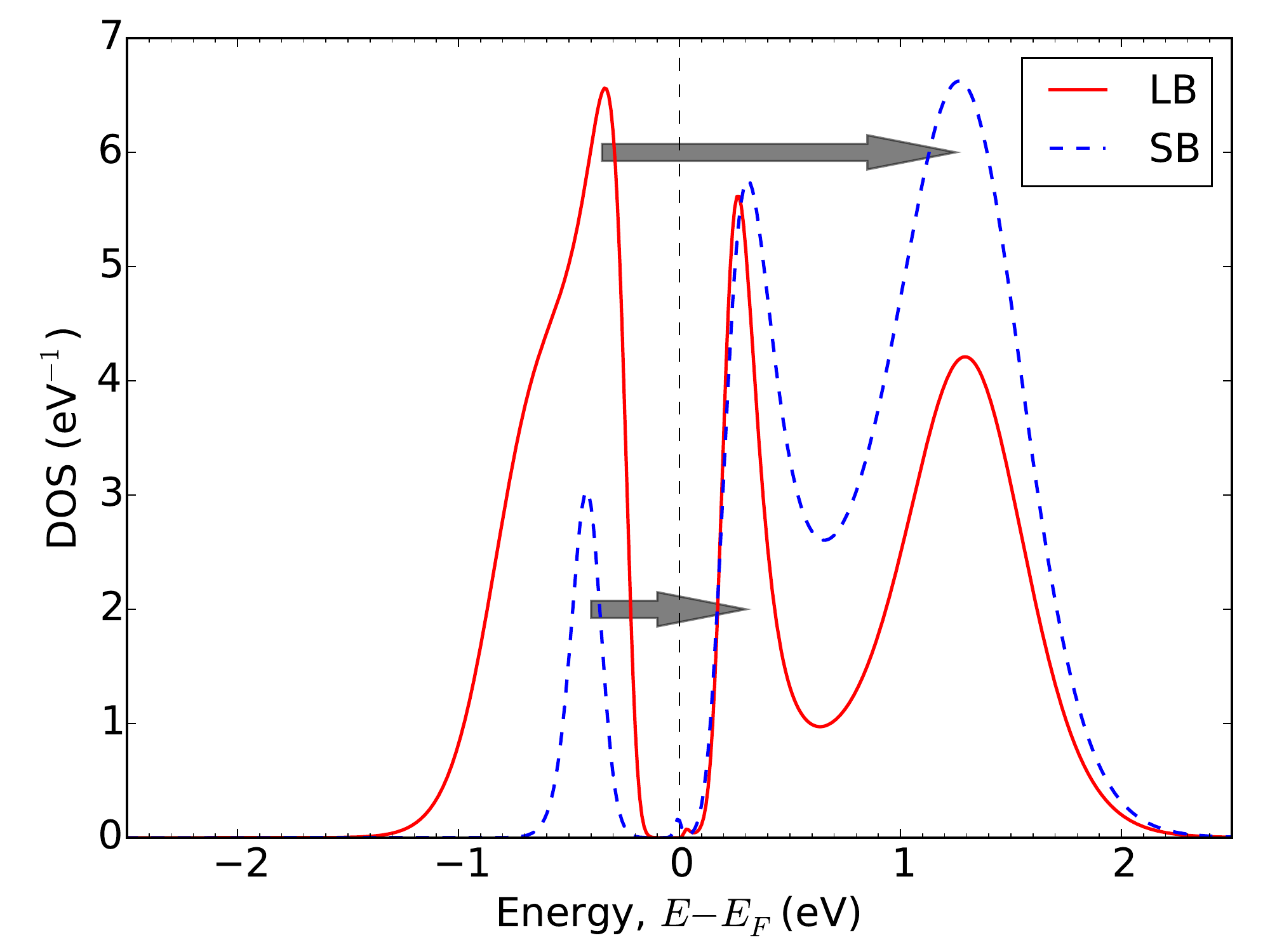}
  \caption{(Color online) Momentum-integrated spectral functions
    (local DOS), as obtained from DMFT (using MaxEnt for analytical
    continuation) for $U=1.0,J=0.8$~eV.  Top: metallic \ortho
    structure.  Bottom: insulating \mono structure (LB: long-bond
    sites, SB: short-bond sites).  Arrows indicate the energies of the
    main optical transitions (see text).
    }
  \label{fig:spectra_U1}
\end{figure}

\begin{figure} 
\includegraphics[width=\columnwidth]{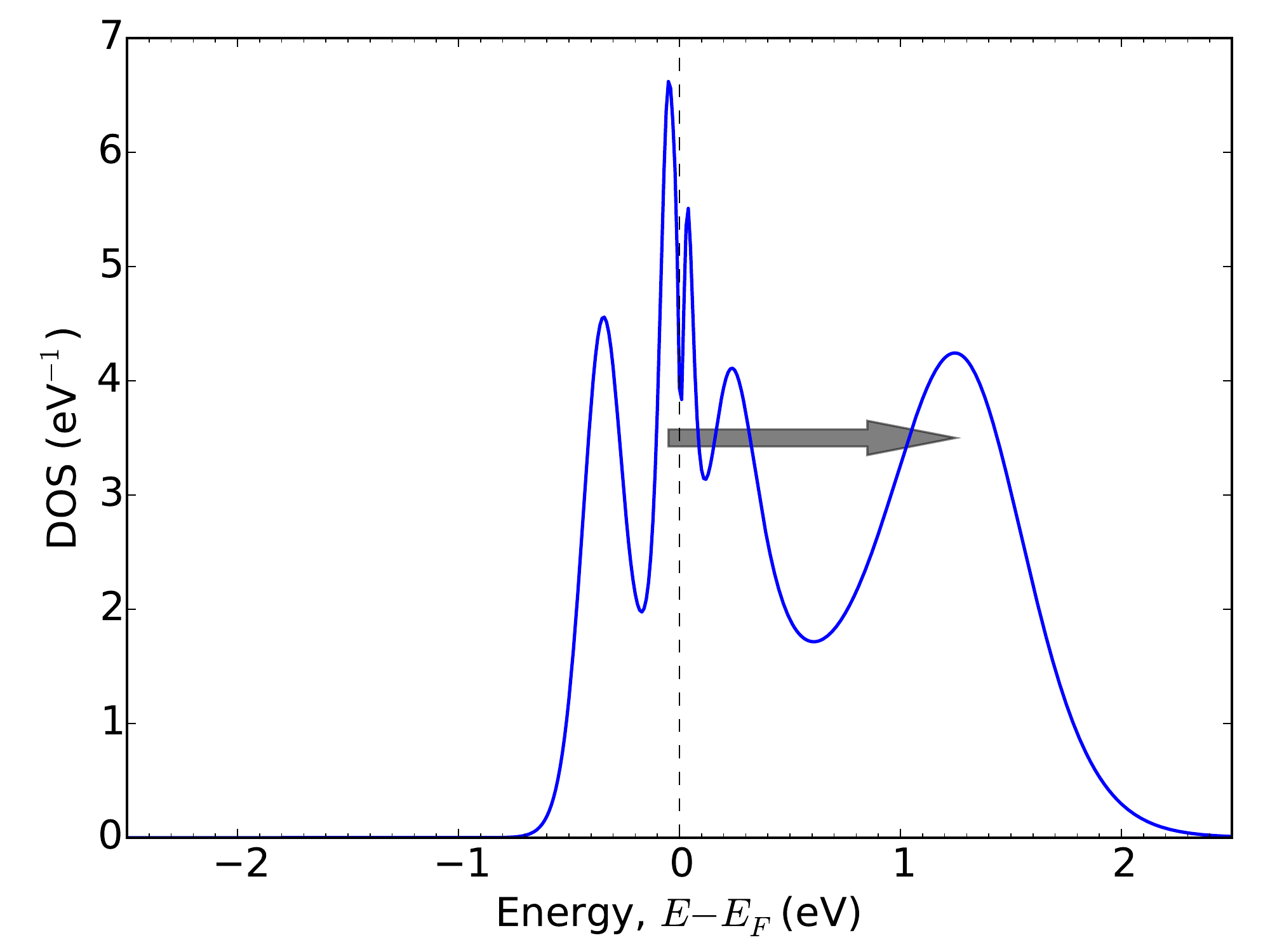}\\
\includegraphics[width=\columnwidth]{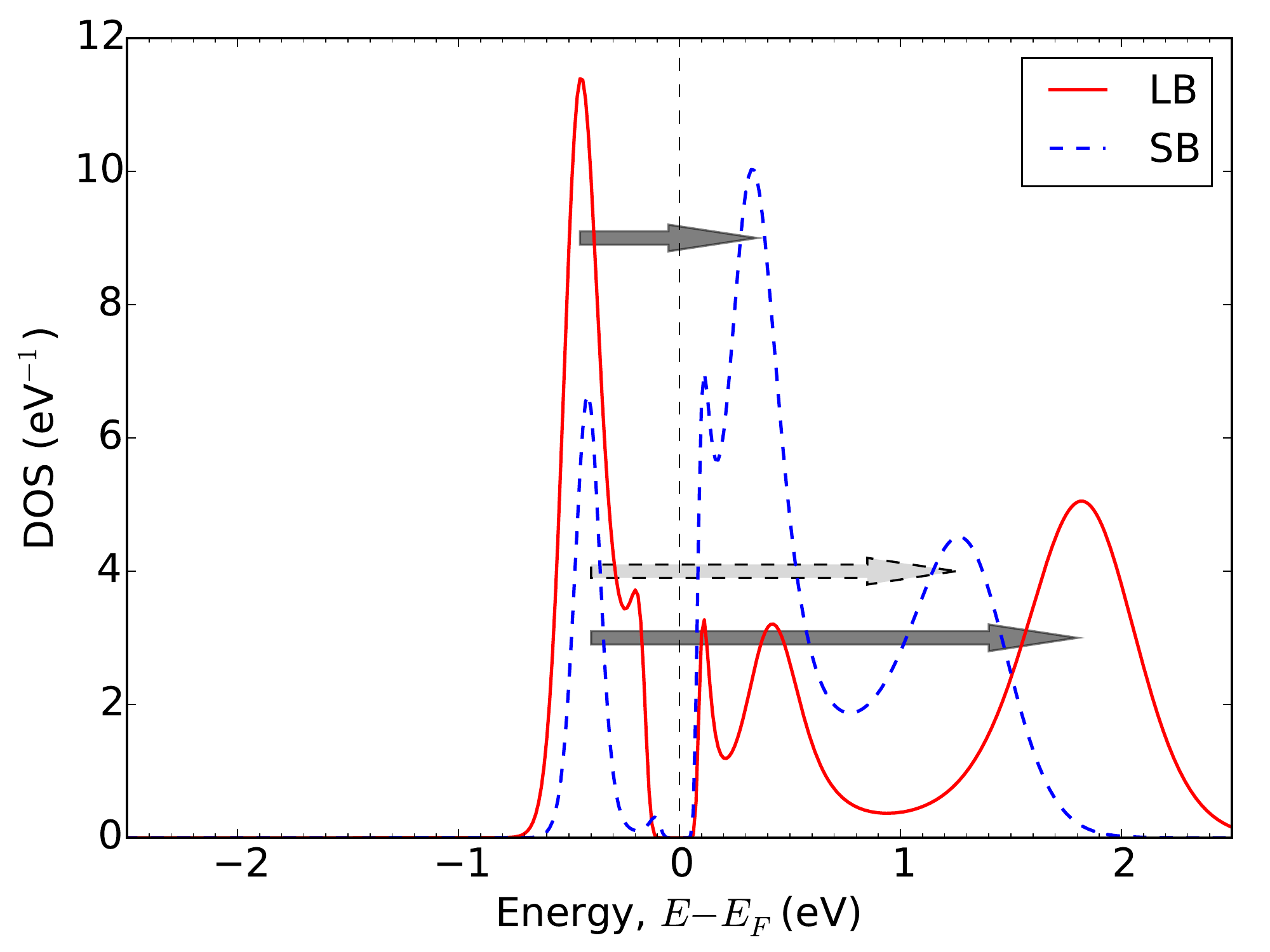}
  \caption{(Color online) Same as Fig.~\ref{fig:spectra_U1}, for
    $U=2.0, J=0.8$~eV.  }
  \label{fig:spectra_U2}
\end{figure}

We display in Fig.~\ref{fig:spectra_U1} and Fig.~\ref{fig:spectra_U2}
the momentum-integrated spectral function (total density of states) of
the metallic state (\ortho structure) and of the insulating state
(\mono structure) for two sets of interaction parameters:
$U=1,J=0.8$~eV and $U=2,J=0.8$~eV.  These results were obtained by
analytically continuing the DMFT Monte-Carlo data to the
real-frequency axis (for details, see Appendix~\ref{sec:appendix}).
These two sets of interaction parameters are illustrative of two
different regimes for the insulating state.  The first one is close to
the lower side of the BDI lobe in Fig.~\ref{fig:phasediag} (bottom
panel), so that the gap is controlled by the proximity to the lower
critical boundary $U+J\simeq W_{<}$.  The second one is close to the
upper critical boundary and the bond-disproportionation crossover line
$U-3J=\delsite$, so that the gap is controlled by the proximity to
this line.

Let us first turn our attention to the spectra of the metallic phase
(Fig.~\ref{fig:spectra_U1} and Fig.~\ref{fig:spectra_U2}, top
panels). They display a peak centered around the Fermi level,
corresponding to quasiparticle excitations.  This quasiparticle DOS is
narrowed down by correlation effects in comparison to the LDA DOS
(Fig.~\ref{fig:dos_LDA}, top panel), the narrowing being less
pronounced for $U=1, J=0.8$ eV than for $U=2, J=0.8$ eV, in line with
the smaller value of the quasiparticle weight $Z$ for the latter
($Z\simeq 0.55$ and $\simeq 0.35$, respectively).  Side peaks within
the lowest (dominantly LB) band, spanning the energy window $\sim
[-0.5,+0.5]$~eV are also apparent.
The most prominent spectral feature aside from the central
quasiparticle peak is the large peak on the positive energy side, at
about $\sim 1.2-1.3$~eV. This peak is separated from the QP peak by a
pronounced dip at $\sim +0.5$~eV.  The energy of this spectral feature
does not depend very much on the value of $U$. Indeed, in
Appendix~\ref{sec:appendix}, we display for completeness the spectra
for $U=3.5,J=1.2$~eV and we see that this high-energy peak is still
well below $2.0$~eV in the metallic phase.  Note that this large value
of the effective $U$ is likely to be too large to describe nickelates
properly.
The dip at $\sim +0.5$~eV and prominent peak above this energy scale
are already apparent in the LDA DOS of Fig.~\ref{fig:dos_LDA}.  The
dip corresponds to the crossing points between the two band manifolds
which are subsequently split-off by the bond disproportionation, and
the high-energy peak corresponds to the empty states which form the
upper manifold.
We observe that the LDA DOS displays another dip at $\sim 1.1$~eV
corresponding to another set of crossing points in the LDA
bandstructure.  It may be that such a dip should also be present in
the presence of correlations, but that the analytical continuation
procedure smears out spectral features and is thus insufficient to
reveal it.

We now turn to the spectra in the bond-disproportionated insulating
state, displayed for each type of sites (LB and SB) in
Fig.~\ref{fig:spectra_U1} and Fig.~\ref{fig:spectra_U2} (bottom
panels, see also Fig.~\ref{fig:spectra_U35} in the Appendix).
Below the insulating gap, the spectra display a lower Hubbard band
(LHB) corresponding to the removal of a single electron. In the
extreme limit of $\eg^2\eg^0$ occupancy, this LHB would be entirely of
LB nature, and no LHB would be visible for SB sites.  Because the
$\eg$ charge imbalance is not complete, both sites display a LHB, but
its spectral intensity is indeed larger for LB sites, as expected.

%
%

The positive energy states, corresponding to the electron addition
spectrum, displays a richer structure.  Both the SB and LB spectra
reveal two marked spectral features on the positive energy side.
Let us designate these two positive-energy spectral peaks for LB sites
by \LBlow and \LBhi, in increasing order of energies (similarly,
\SBlow, \SBhi).
By comparing Figs.~\ref{fig:spectra_U1}, \ref{fig:spectra_U2} and
\ref{fig:spectra_U35}, we observe that the position of the peaks
\LBlow and \SBlow are always similar. In contrast, the energy of \LBhi
and \SBhi are different in general. The energy of \SBhi is around
$\sim 1.3$~eV, with little dependence on $U$.  At the same time, the energy
of \LBhi clearly increases with $U$.
Coincidentally, the energy of \LBhi and \SBhi are approximately equal
for the parameter set $U=1,J=0.8$~eV, but this does not hold true for
the other cases.

\subsubsection{Interpretation of the spectral features}

A simple analysis in the atomic limit allows us to understand the nature of
these spectral features, starting from the extreme limit in which the ground
state is $\eg^2\eg^0$.  
Removing an electron on a LB site corresponds to a transition energy between the final and initial states:
\begin{eqnarray}
\nonumber
\Delta E_{\mathrm{LHB}}
& = & -\frac{\delsite}{2}-\mu -
\left[ -\delsite-2\mu+U-3J\right]\\
& = & \mu+\frac{\delsite}{2}-(U-3J),
\end{eqnarray}
which sets the position of the lower Hubbard band with respect to the
Fermi level $\mu$.  Similarly, adding an electron on a LB site (upper
Hubbard band for LB sites) corresponds to:
\begin{eqnarray}
\nonumber
\Delta E_{\mathrm{UHB}}^{\mathrm{LB}}
& = & -3 \frac{\delsite}{2}-3\mu +U+(U-2J)+(U-3J) 
\\ \nonumber
& & - \left[ -\delsite-2\mu+U-3J\right]
\\
& = &  -\mu-\frac{\delsite}{2}+2U-2J,
\end{eqnarray}
and adding an electron to an SB site yields: 
\begin{equation}
\nonumber
\Delta E_{\mathrm{UHB}}^{\mathrm{SB}}\,=\,\frac{\delsite}{2}-\mu .
\end{equation}
Hence, the energy separation between the LHB and the upper Hubbard
band (electron addition peak) corresponding to a LB site reads, in the
atomic limit:
\begin{equation}
\Delta E_{\mathrm{LHB}}+\Delta E_{\mathrm{UHB}}^{\mathrm{LB}}\,=\,U+J
\label{eq:UHB_LB}
\end{equation} 
while, for a SB site:
\begin{equation}
\Delta E_{\mathrm{LHB}}+\Delta
E_{\mathrm{UHB}}^{\mathrm{SB}}\,=\,\delsite-(U-3J) .
\label{eq:UHB_SB}
\end{equation} 
Not surprisingly, we recover the two key energy scales discussed
above: $U+J$ corresponds to the effective $U$ for a half-filled
orbital and controls (when compared to $W_{<}$) the lower critical
boundary of the BDI state, while $\delsite-(U-3J)$ is the energy scale
controlling bond disproportionation which sets the disproportionation
crossover line as well as the upper critical boundary of the BDI
state.

These considerations allow us to unambiguously identify the highest of
the two LB spectral features, \LBhi, as the corresponding upper
Hubbard band. Indeed its separation from the LHB is quite well
approximated by $\sim U+J$ (yielding $1.8$, $2.8$ and $4.7$~eV for
Fig.~\ref{fig:spectra_U1}, Fig.~\ref{fig:spectra_U2} and
Fig.~\ref{fig:spectra_U35}, respectively).  This separation is
indicated by the longest plain grey arrow on these figures.
The above estimate $\delsite-(U-3J)$ for the UHB of SB sites is also
in reasonable agreement with the electron addition peak of highest
intensity on SB sites, with an estimated value $1.65$, $0.65$ and
$0.35$~eV for Fig.~\ref{fig:spectra_U1}, Fig.~\ref{fig:spectra_U2} and
Fig.~\ref{fig:spectra_U35}, respectively (also depicted as a plain
grey arrow on these figures).  We observe that this highest-intensity
peak corresponds to the \SBhi feature for the $U=1,J=0.8$ eV case
(Fig.~\ref{fig:spectra_U1}), but to the \SBlow feature for the two
other cases.
The existence of a second peak (besides the UHB) in the SB spectra,
separated by a dip from the other SB spectral peak is, in our opinion,
due to the structure of the DOS of the SB upper band manifold already
visible at the LDA level.  Indeed, we observe that the separation
between the two SB spectral features is always of order $\sim 1.0$~eV,
which is of a magnitude similar to the separation of the two peaks of
the upper manifold in the LDA DOS.
%

The fact that $U+J=1.8$~eV and $\delsite-(U-3J)\simeq 1.65$~eV are
approximately equal for $U=1.0, J=0.8$ eV explains that the \LBhi and \SBhi
(UHB) features are, coincidentally, located at approximately equal
energies in this specific case.  The condition for such a coincidental
overlap of the two LB- and SB- UHB is, in the atomic limit from
Eqs.~(\ref{eq:UHB_LB},\ref{eq:UHB_SB}): $U=J+\delsite/2$. For
$U>J+\delsite/2$, the LB-UHB lies above the SB-UHB, while the opposite
holds true when $U<J+\delsite/2$.

\subsubsection{Consequences for optical spectroscopy} 

These considerations have direct consequences for optical spectroscopy, which can be anticipated by 
comparing the top and bottom panels of Fig.~\ref{fig:spectra_U1} and Fig.~\ref{fig:spectra_U2} in order 
to understand how the optical spectra of nickelates change through the MIT. We only provide a qualitative 
discussion here, leaving a detailed calculation of optical spectra and comparison to experiments for future 
work.

Optical spectroscopy involves particle-hole transitions, hence energy differences between the main spectral 
features of the one-particle spectral function displayed in these figures. Of course, these transitions are weighted 
by matrix elements of the current, which require a detailed calculation of the band velocities (transport function). 
The current is an inter-site process, which will dominantly couple LB to SB sites, so that we should direct our attention 
to the energy differences between LB and SB spectral features in the one-particle spectra. 

In the metallic phase, we expect the optical conductivity to consist
of two main spectral features.  A Drude peak at low-energy (with
reduced weight $\sim Z$) involving near Fermi level transitions, and
an additional feature at $\sim 1.2$~eV (indicated by the plain arrow
on the top panels of the figures, and rather independent of $U,J$),
corresponding to the optical transitions between the states near Fermi
level and the states forming the upper band submanifold, as discussed
above.

In the insulating state, we expect generically three spectral features
above the absorption edge set by the gap.  The lowest-energy one
corresponds to transitions from the LHB to the \LBlow/\SBlow lowest
spectral feature (indicated by the shortest arrow in the bottom panel
of the figures).
The highest-energy one corresponds to transitions from the LHB to the highest of the 
two high-energy spectral features (i.e. for $U>J+\delsite/2$, the LB-UHB, \LBhi excitation), 
indicated by the longest arrow on the figures.
An intermediate energy optical feature corresponds to the transition between the LHB and the 
lowest of the two high-energy spectral features (dashed arrow on the figures). 
As discussed above, the intermediate and high-energy optical transition merge  
when $U\sim J+\delsite/2$ (as in Fig.~\ref{fig:spectra_U1}), leaving only two peaks 
above the absorption edge. 

Hence, the distinctive signature of the MIT is expected to be the
splitting of the $\sim 1.2$~eV peak in the metallic phase into two (or
three) peaks, one of lower and one of higher energy
(Figs.~\ref{fig:spectra_U1} and \ref{fig:spectra_U2}).  Indeed, such a
splitting has been observed in recent optical spectroscopy experiments
\cite{Ruppen2014}. We propose that our low-energy theory provides a
theoretical interpretation of this experimental discovery, and hope to
support this claim by explicit calculations of optical spectra in
future work.

\subsection{Local magnetic moments, NMR and relation to charge disproportionation}
\label{sec:mag}

\begin{figure} 
\includegraphics[width=\columnwidth]{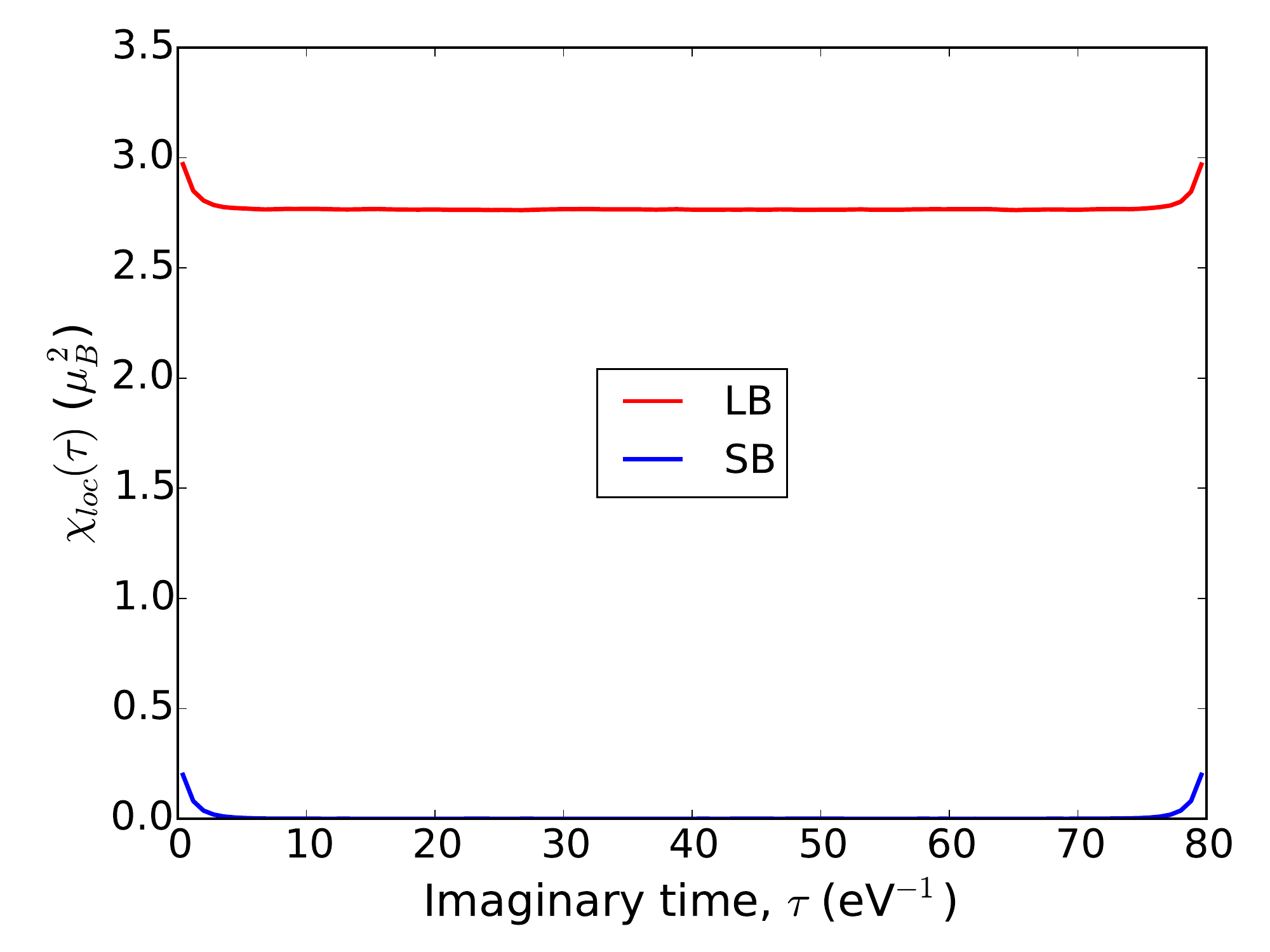}\\
\includegraphics[width=\columnwidth]{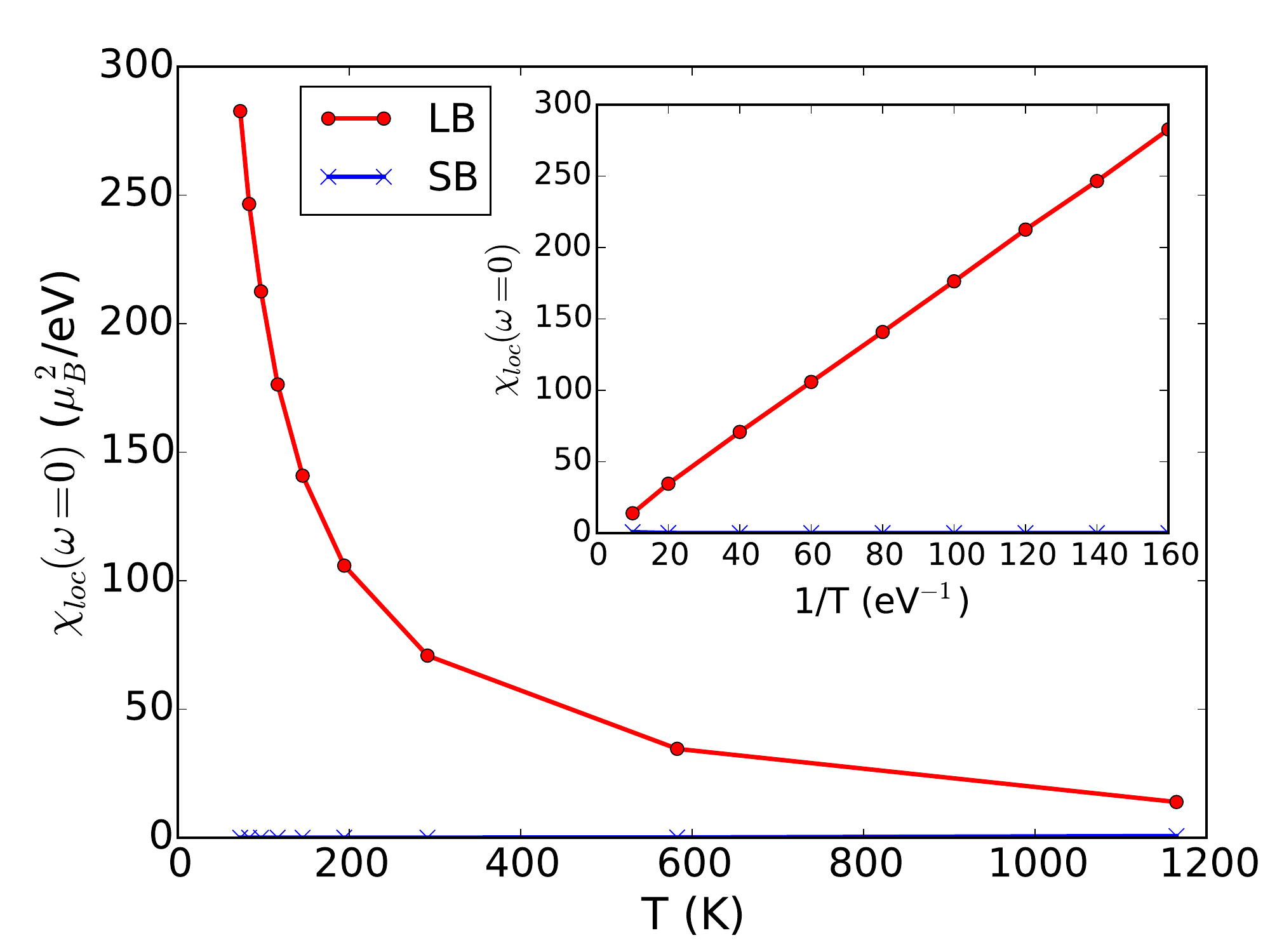}
  \caption{(Color online) Local spin-spin correlations of Ni-LB and
    Ni-SB sites in the insulating \mono structure, for $U=1 ,
    J=0.8$~eV.  Top: Local correlation function
    $\chi_{\mathrm{loc}}(\tau)=\langle S^z(0)S^z(\tau)\rangle$ as a
    function of imaginary time, at a temperature $T=1/80$~eV$\simeq
    150$~K.  Bottom: Static zero-frequency value
    $\chi_{\mathrm{loc}}(\omega=0)$ as a function of temperature.  }
  \label{fig:chiloc}
\end{figure}

We now discuss the consequences of the bond disproportionation for the local magnetic properties 
of the SB and LB sites in the paramagnetic insulating phase (BDI). 
As pointed out in Refs.~\cite{Mizokawa2000,Mazin2007,Park2012,Johnston2014}, 
in the extreme $d^8\Lbar^2$/$d^8$ picture, one expects the SB octahedra to carry
no magnetic moment and the LB ones to carry a full $S=1$ moment.
Indeed, in this picturue the spin-$1$ moment (corresponding to two
$\eg$ electrons with parallel spins in different orbitals in the Ni
$d^{8}$ configuration) on SB octahedra is screened by the two ligand
holes, while no such screening takes place on the LB sites. The
analogy to a mixed Kondo-insulator (SB sites)/Mott insulator (LB
sites) state was emphasized in Ref.~\cite{Park2012}.

Reproducing such a behaviour is a challenge to any low energy description that does not include 
explicitly the oxygen states. Here, we show that our low-energy two-orbital model does achieve this 
goal in the regime $U-3J<\delsite$.

To support this claim, we have calculated the local spin-spin
correlation on both the LB and SB sites.  In Fig.~\ref{fig:chiloc}
(top panel), we display this correlation function
$\chi_{\mathrm{loc}}(\tau)=\langle S^z(0)S^z(\tau)\rangle$ as a
function of imaginary-time in the BDI phase at a temperature
$T=1/80$~eV$\simeq 150$~K. In the bottom panel of the same figure, we
display the zero-frequency value of the local susceptibility
$\chi_{\mathrm{loc}}(\omega=0,T)$ as a function of temperature.
These data clearly signal the presence of a fluctuating local moment
on LB sites: $\chi_{\mathrm{loc}}(\tau)$ does not decay at large
$\tau$ ($\sim \beta/2$), and $\chi_{\mathrm{loc}}(\omega=0,T)$ follows
a Curie law $\sim 1/T$.  The corresponding value of the local moment
obtained from a fit to $\chi = \mu_B^2
S_{\textrm{eff}}(S_{\textrm{eff}}+1)/3T$ is
$S_{\textrm{eff}} = 1.86$  $\mu_{B}$.
In contrast, on SB sites $\chi_{\mathrm{loc}}(\tau)$ and
$\chi_{\mathrm{loc}}(\omega=0,T)$ have a very small
temperature-independent value, signalling the absence of a local
moment.

Obviously, the detection of a clear separation between moment-carrying
LB octahedra and non-magnetic (or at least reduced moment) SB
octahedra would be a smoking gun for the physical picture proposed in
earlier works and for the low-energy description introduced here.
Further studies of the insulating phase of nickelates in the range
$T_N<T<T_{\mathrm{MIT}}$ using local magnetic probes such as NMR or
muon-diffraction, or inelastic neutron scattering is therefore highly
desirable.

\begin{figure} 
\includegraphics[width=\columnwidth]{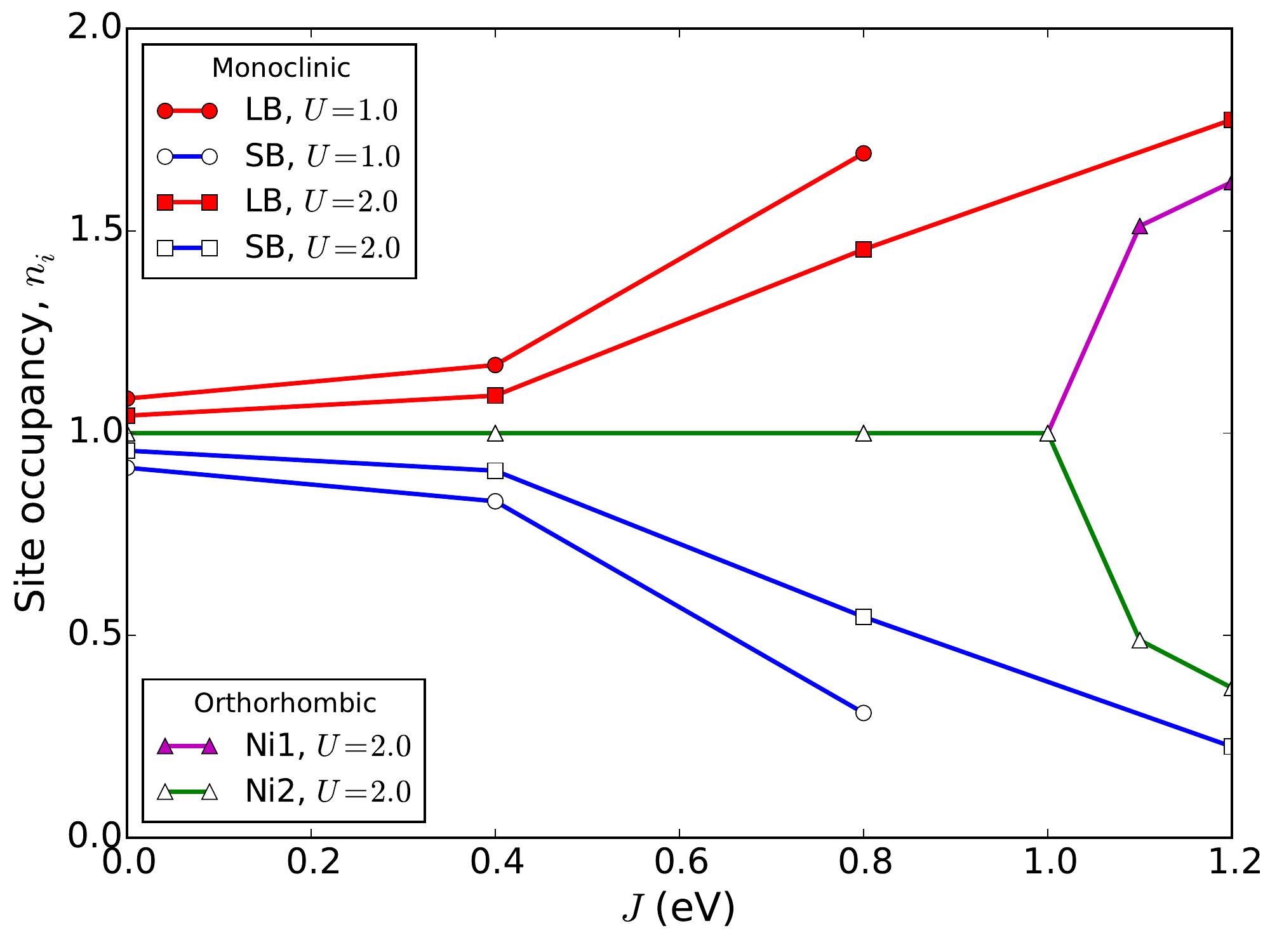}
  \caption{(Color online) Occupancies of LB and SB sites as a function
    of $J$ for $U = 1.0$ and $2.0$ eV.  Also shown are the occupancies
    (triangles) of the two types of Ni sites in the orthorhombic
    structure experiencing a spontaneous symmetry breaking at $J\sim
    1.0$ eV.}
  \label{fig:ndisp}
\end{figure}

Finally, we show the evolution of the LB and SB site occupancies as a
function of $J$ in Fig.~\ref{fig:ndisp}. 
The $e_{g}$ charge disproportionation is clearly enhanced as one
approaches the BDI phase. In contrast, it is suppressed as one
moves towards the Mott phase, as indicated by a smaller difference of
the site occupancies for the larger value of $U$. 
It is important to note that the $\eg$
occupancies should not be interpreted as indicative of the valence of nickel atoms  
because the $\eg$ orbitals are antibonding linear combination of
the Ni $d$ and O $p$ orbitals. As the $\eg$ states have substantial O
$p$ character, the disproportionation of the $\eg$ occupancy could
primarily be the result of a change in occupation of the O $p$
states (ligand holes). Some disproportionation of the atomic Ni charge cannot be
ruled out within our low energy model, however. 

Furthermore, the charge of an ion in a solid compound with strongly
covalent bonds is not a well-defined quantity.  This implies that
x-ray or neutron scattering experiments do not 
distinguish whether the disproportionation takes place within the
atomic-like Ni $d^7$ orbitals, $d^8\Lbar$ states in the small or
negative charge transfer regime, or the antibonding $\eg$ states.
Probes such as polarized soft x-ray absorption
(XAS) or resonant inelastic x-ray scattering (RIXS) spectroscopy are actually 
sensitive to the occupancy and symmetry of the antibonding valence $\eg$ states, 
since in those spectroscpies an electron is transferred from a core state into such 
an empty $\eg$ state. Therefore, we believe that available experimental results~\cite{Bodenthin2011, Wu2013} 
usually interpreted as `charge' disproportionation or `charge' ordering are actually 
consistent with a disproportionation of the $\eg$ occupancies. 

In our low-energy description of the nickelates, the bond
disproportionation plays an essential role in the electronic
properties. This strong coupling between the lattice and electronic
properties should also manifest in the phonon properties. In
particular, the zone-center Raman mode corresponding to the expansion
and compression, respectively, of alternate sets of corner-shared
NiO$_6$ octahedra of the orthorhombic phase should show a large and
increasing linewidth as the temperature is lowered down to $\Tmit$. In
addition, this phonon mode should soften and show an unstable behavior
upon reaching $\Tmit$.

\section{Conclusion and Perspectives}
\label{sec:conclusion}

We have used combined DFT+DMFT calculations to construct and study in
detail a low-energy model for the rare-earth nickelates that describes
the metal-insulator and orthorhombic-monoclinic structural transitions
observed in these materials. The DFT calculations show that the $\eg$
manifold is well separated and quarter-filled in the high-temperature
orthorhombic phase.  The major effect of the bond disproportionation in
the low-temperature monoclinic phase is to split the $\eg$ manifold
into two sub-manifolds. This reduces the average degeneracy and the
effective bandwidth. We capture these features of the electronic
structure by constructing a simple model with only two $\eg$ orbitals
per Ni site and a parameter $\delsite$ that describes the difference
in energy between the $\eg$ states at the two bond-disproportionated
Ni sub-lattices. Although we explicitly construct this model for
LuNiO$_3$, our model is generic and applicable to other members of the
rare-earth nickelates family.

We solve this model using DMFT calculations for a wide range of values
of the on-site Coulomb repulsion $U$ and Hund's rule coupling $J$ for
the physically relevant small values of $\delsite$, as well as for
larger values.  We find that the physics of nickelates can be
consistently accounted for in the regime where $U-3J\leq\delsite$ is
small or negative, without a fine tuning of the interaction
parameters.
In particular, a bond-disproportionated paramagnetic insulating state
is stabilized for a wide range of interaction parameters in this
regime.
Furthermore, we find that in this regime and for large
enough (but realistic) values of $J>J_c$, the metallic state is
spontaneously unstable to disproportionation.  This rationalizes the
large sensitivity to bond-length disproportionation observed in these
materials.

Our minimal theory demonstrates that the MIT of nickelates cannot be viewed as the 
Mott transition of a homogeneously quarter-filled band. Furthermore, it emphasizes 
that a small or negative charge transfer energy, a sizeable Hund's coupling and a strong 
coupling to lattice effects through bond-length disproportionation are 
essential to the physics, hence confirming and unifying the proposals of 
previous authors~\cite{Mizokawa2000,Mazin2007,Park2012,Johnston2014}. 

The main advantages of the minimal low-energy description proposed here is 
its universality: it can easily be applied to other nickelates and other materials with 
similar physics. Quantitative calculations using DMFT for this low-energy model are 
considerably simpler than a full-fledge DFT+DMFT treatment involving both 
transition-metal and ligand states, and are free of the ambiguities associated with double-counting. 
The one-particle part of the low-energy 
Hamiltonian can be easily adapted to the specific material of interest using standard 
electronic structure and Wannier functions techniques. Furthermore, as 
demonstrated in this article, the main aspects of the phase diagram and underlying 
physics of this low-energy model can be rationalized and explained 
using simple qualitative arguments. 

In contrast, the coupling constants $U$ and $J$ entering our theory are low-energy parameters 
that we did not attempt to derive from first principles. We would like to propose this 
as an important physical test of first-principle methods aiming at the determination 
of such low-energy parameters. Whether these methods can access the regime of small 
or negative charge transfer and whether they will support the view that nickelates (and 
other materials with similar physics) belong to this regime is an outstanding challenge.

Finally, we have outlined some experimental implications of the low-energy effective 
theory proposed in this paper. Some of these appear to be consistent 
with very recent experimental findings from optical spectroscopy~\cite{Ruppen2014}. 
More theoretical and experimental work is obviously needed however to put our low-energy 
description to the test.

\begin{acknowledgments}
We are grateful to Michel Ferrero for discussions and help with DMFT calculations, and 
to Dirk van der Marel for sharing with us his optical data prior to publication. 
We also acknowledge useful discussions with Sara Catalano, Stefano Gariglio, Marta Gibert, Jean-Marc Triscone and 
all other members of the Triscone group in Geneva, as well as with Leon Balents, Philipp Hansmann, Andrew J. Millis, 
Leonid Pourovskii, 
George Sawatzky and Michel Viret. 
Support for this work was provided by the Swiss National Science Foundation
(grant 20021-146586 and NCCR-MARVEL),  
by a grant from the European Research Council (ERC-319286 QMAC), 
and by the Swiss National Supercomputing Centre (CSCS) under project ID s497. 
\end{acknowledgments}

\appendix
\section{Methods and technical information on the calculations}
\label{sec:appendix}

\subsection{Electronic structure calculations}

The electronic structure calculations presented in this article were
obtained within the local density aproximation (LDA) using the general
full-potential linearized augmented planewave method as implemented in
the WIEN2k software package \cite{wien2k}. Muffin-tin radii of $2.28$,
$1.91$, and $1.64$~a.u. for Lu, Ni, and O, respectively, were used.
An $8\times 8 \times 8$ $\mathbf{k}$-point grid was used to perform
the Brillouin zone integration in the self-consistent
calculations. The planewave cutoff was set by $RK_{\mathrm{max}} = 7$,
where $K_{\mathrm{max}}$ is the planewave cutoff and $R$ is the
smallest muffin-tin radius used in the calculations.
The calculations were performed at experimental values of the lattice parameters and 
atomic positions, as given for \lunio in Ref.~\cite{Alonso2001} at a temperature of 
$673$~K for the orthorhombic structure and of $533$~K for the monoclinic structure. 
In the orthorhombic structure, the Ni-O-Ni angles are 145$^\circ$ and 143$^\circ$ along
the $ab$ plane and $c$ axis, respectively (they would be 180$^\circ$ in the hypothetical 
undistorted cubic structure). 
Similar values of the angles hold in the \mono structure. 
Such relatively large rotations induce a clear separation between the manifold of the Ni-$\eg$bands 
and that of the Ni-$t_{2g}$ ones and significantly reduce the bandwidth of the $e_g$ manifold 
(by approximately a factor of two, as compared to the cubic structure).  
%


\subsection{DMFT calculations}

\begin{figure} 
\includegraphics[width=\columnwidth]{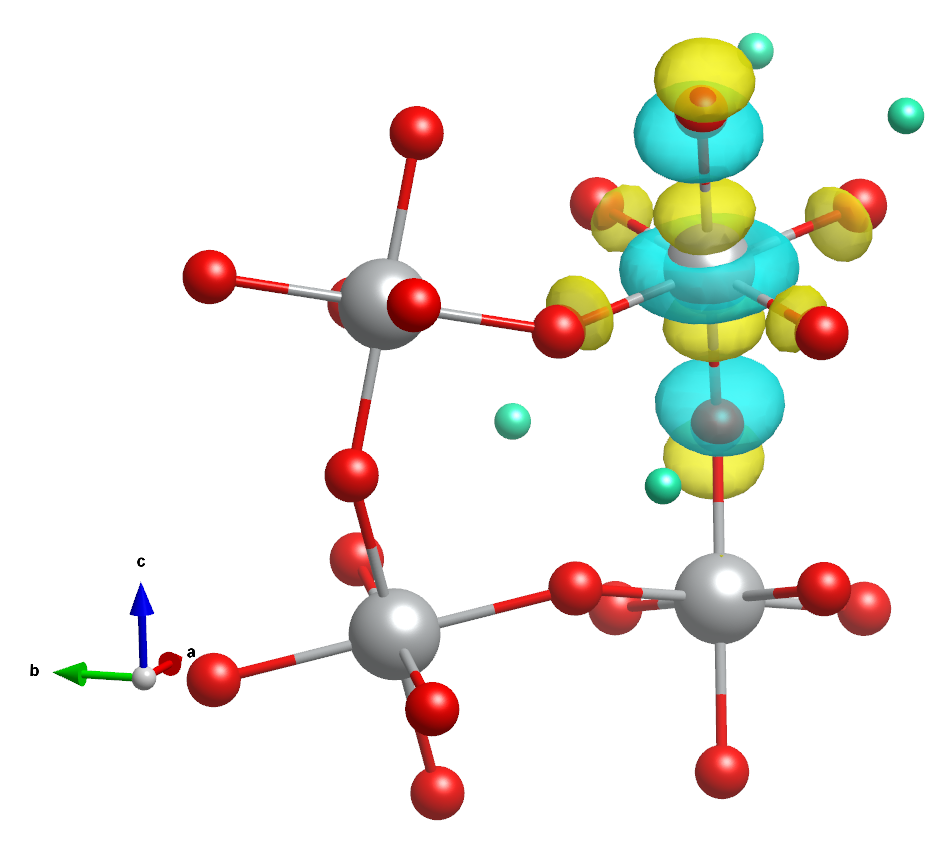}\\
\includegraphics[width=\columnwidth]{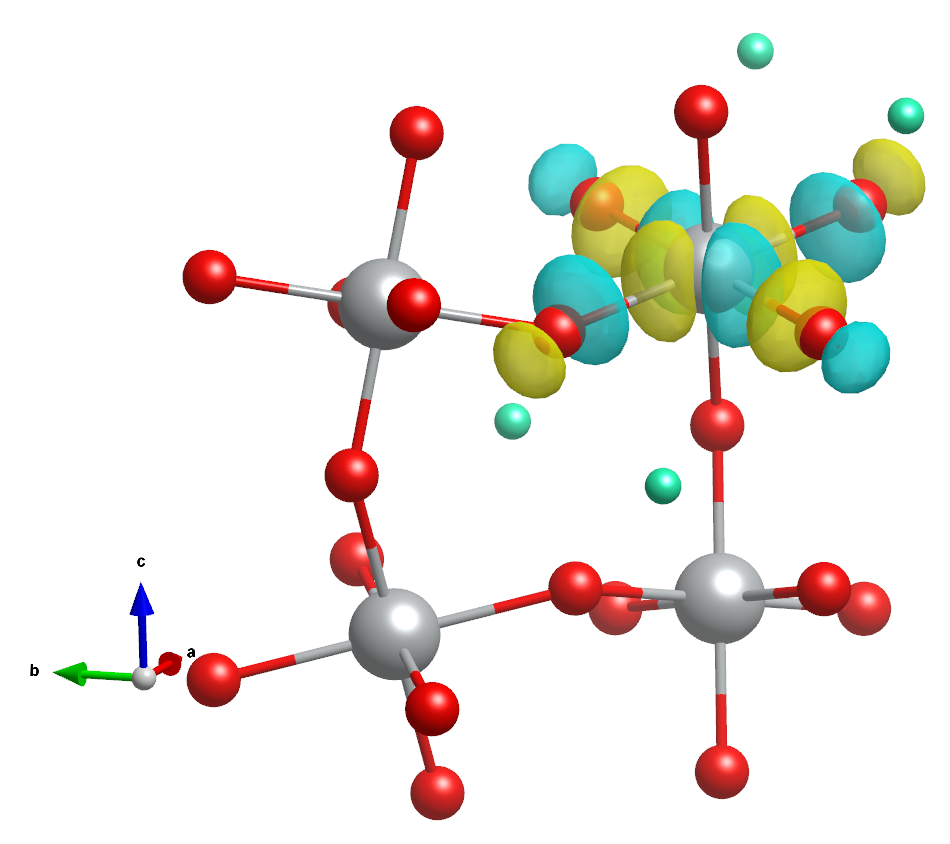}
  \caption{(Color online) Wannier functions corresponding to PLOs for
    Ni $e_{g}$ orbitals for one of the Ni ions (large grey balls). The
    Wannier functions are very similar in both the \mono (for two
    types of Ni sites) and \ortho structures.  Note that a significant
    contribution is coming from oxygen ions (medium-size red
    balls). The size of La ions (small cyan balls) is reduced for a
    better presentation. Plotted using VESTA \cite{VESTA}.}
  \label{fig:wann}
\end{figure}

All DMFT calculations presented in this paper have been performed
using the continuous-time hybridization expansion Monte Carlo solver~\cite{Gull2011}
as implemented in the TRIQS software library~\cite{TRIQS}.
The local Green's function have been sampled in Legendre basis
\cite{Boehnke2011} and analytical continuation has been performed
using a maximum entropy (MaxEnt) method.
Combined DFT+DMFT calculations have been done by making use of
two different interfaces: one based on maximally localized Wannier 
functions (MLWF) \cite{Marzari1997,Wannier90,Marzari2012_review}
constructed from the WIEN2k band structure and another one
based on projector localized orbitals (PLO) \cite{Amadon2008}
obtained from the Vienna \textit{ab-initio}
simulation package (VASP) \cite{paw_vasp,vasp1,vasp2}
(VASP band structure have been calculated for the same structure parameters
as given in the previous subsection, with the k-mesh containing
$10\times10\times7$ points and the plane-wave cutoff being
$E_{\textrm{cut}} = 500$ eV). Examples of the resulting Wannier functions
are presented in Fig.~\ref{fig:wann}. Both approaches, MLWF and PLO, lead to almost identical
outcomes (including the band structure) and we have thus checked that 
our results and conclusions are not sensitive to a particular choice of the DFT+DMFT method.

%
%

For MLWFs (PLOs) we have used an energy window of $[-0.4, +2.0]$~eV ($[-0.6, +2.6]$~eV)
which encloses the eight bands of the $e_g$ manifold, 
and obtain two Wannier functions per each Ni site, corresponding to the two $e_g$-like orbitals.
The spatial spreads of the MLWFs which were minimized using Wannier90
\cite{Wannier90} came out to be  
Ni$_{LB1}$: 4.32, Ni$_{LB2}$: 4.42,
Ni$_{SB1}$: 4.17, Ni$_{SB2}$: 4.17 (in \AA$^2$).

%
%
\begin{figure} 
\includegraphics[width=\columnwidth]{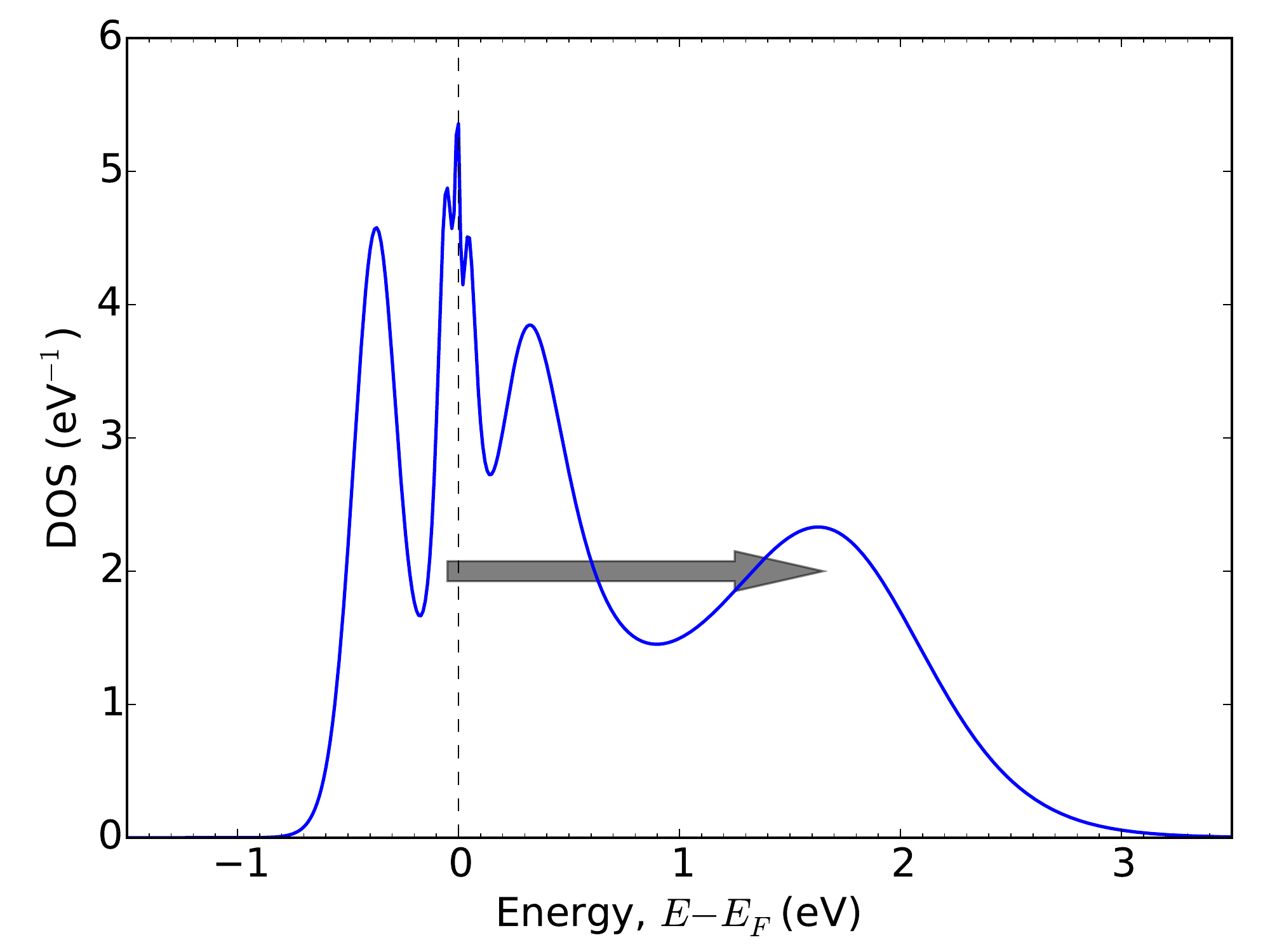}\\
\includegraphics[width=\columnwidth]{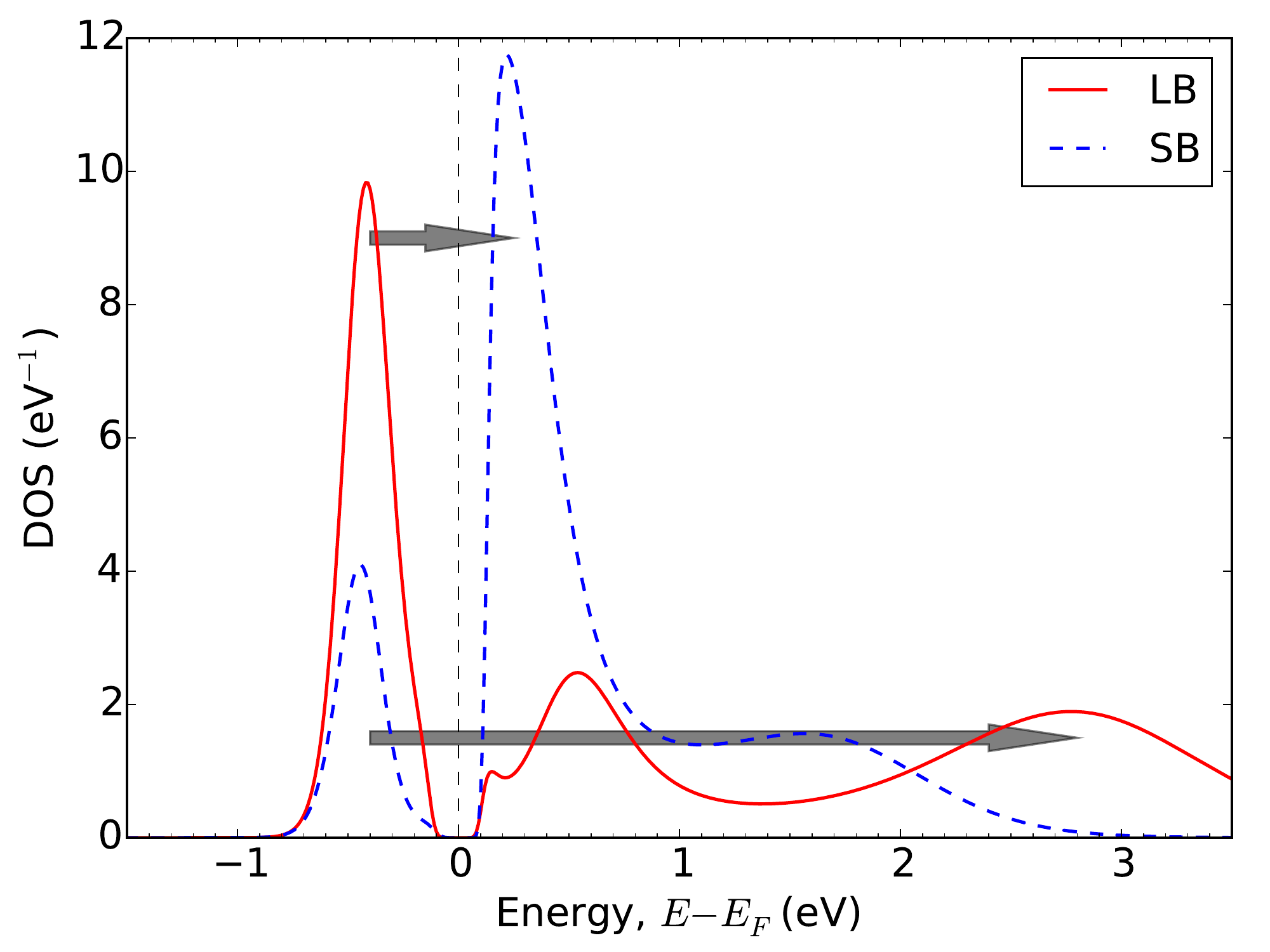}
  \caption{(Color online) Momentum-integrated spectral functions
    (local DOS), as obtained from DMFT for $U=3.5,J=1.2$~eV. Top
    panel: metallic \ortho structure.  Bottom panel: insulating \mono
    structure.
    }
  \label{fig:spectra_U35}
\end{figure}

\section{Qualitative analysis of the two-sublattice two-orbital model}
\label{appendix:model}

We start from a Hamiltonian for a two-sublattice model with
nearest- (NN), $t$, and next-to-nearest-neighbor (NNN), $t'$, hoppings,
\begin{align}\nonumber
H = & -t \sum_{m\sigma\langle i j\rangle} (d^{\dagger}_{m\sigma i} d_{m\sigma j} + h. c.)
 -t' \sum_{m\sigma[i j]} (d^{\dagger}_{m\sigma i} d_{m\sigma j} + h. c.) \\
 & - \frac{\delsite}{2} \sum_{m\sigma i \in A} d^{\dagger}_{m\sigma i} d_{m\sigma i} +
 \frac{\delsite}{2}\sum_{m\sigma j \in B} d^{\dagger}_{m\sigma j} d_{m\sigma j} + 
H_{\mathrm{int}},
\label{eq:ham_model_full}
\end{align}
where the first summation is over pairs of NN sites $i\in A$, $j \in B$, belonging
to the two different sublattices $A$ and $B$, while the second summation is over
pairs of next-to-nearest-neighbor (nnn) sites $i, j \in A$ or $B$ belonging to the same sublattice. 
This Hamiltonian is slightly more general than Eq.~(\ref{eq:ham_model}) because
it also includes the nnn hopping, introduced here to have a more physically meaningful
limit $\delsite \to \infty$; we always assume that $t' \ll t$.
The bare bandwidth can, thus, be still estimated (for a Bethe lattice) as $W_{0} \simeq 4 t$ for $\delsite = 0$.

Consider first the case of $J = 0$. In the limit of large $\delsite$ the model can be viewed as
a half-filled lattice of doubly occupied LB-sites connected via indirect hopping mediated
by empty SB-sites. At $U = 0$, the effective hopping is simply 
$\teff = t' + t^{2} / \delsite$ (provided that $t \ll \delsite$), and the
system is always metallic (with a small bandwidth). 
At finite $U < \delsite$ the energy of the doubly
occupied LB-sites is lifted by $U$ and the indirect hopping is now given by
\begin{equation}
\teff = t' + \frac{t^{2}}{\delsite - U},
\label{eq:teff}
\end{equation}
with the corresponding bandwidth $\Weff = 4 \teff$.

At some critical value $U_{c} = \alpha \Weff$ the LB-sublattice will
experience a Mott transition. For $U \ll \delsite$, this condition
leads to a critical value $U_{c} \simeq c t' + \alpha 4 t^{2} / \delsite$.
More generally, setting $U = U_{c}$ in Eq.~\eqref{eq:teff} (and implicitly
extending the range of validity of this formula to $U \lesssim \delsite$) we get 
a quadratic equation for $U_{c}$,
\begin{align}
U_{c}^{2} - (\delsite + 4\alpha t') U_{c} + \alpha 4 (t^{2} + \delsite t') = 0
\end{align}
leading to two solutions
\begin{align}
U_{c} = \frac{1}{2}\left(\delsite + W' \pm \sqrt{(\delsite - W')^{2} - W^{2}}\right),
\end{align}
with $W' = 4\alpha t'$, $W^{2} = \alpha 16 t^{2}$,
provided that $\delsite > \delc = W + W'$.
When $\delsite < \delc$, the equation has no real solutions implying that
the system is metallic in the entire region $U < \delsite$.

In the limit $\delsite \to \infty$ we can estimate the two critical points as
\begin{align}
U_{c_{1}} \simeq & W' + \frac{W^{2}}{4 \delsite}, 
\label{eq:uc1}\\
U_{c_{2}} \simeq & \delsite - \frac{W^{2}}{4 \delsite},
\label{eq:uc2}
\end{align}
which emphasizes the physical meaning of the asymptotic phase boundary
$U = \delsite$ and also shows that in the limit $\delsite = \infty$
the lower critical value is entirely determined by the nnn hopping.

%
%
\begin{figure} 
\includegraphics[width=\columnwidth]{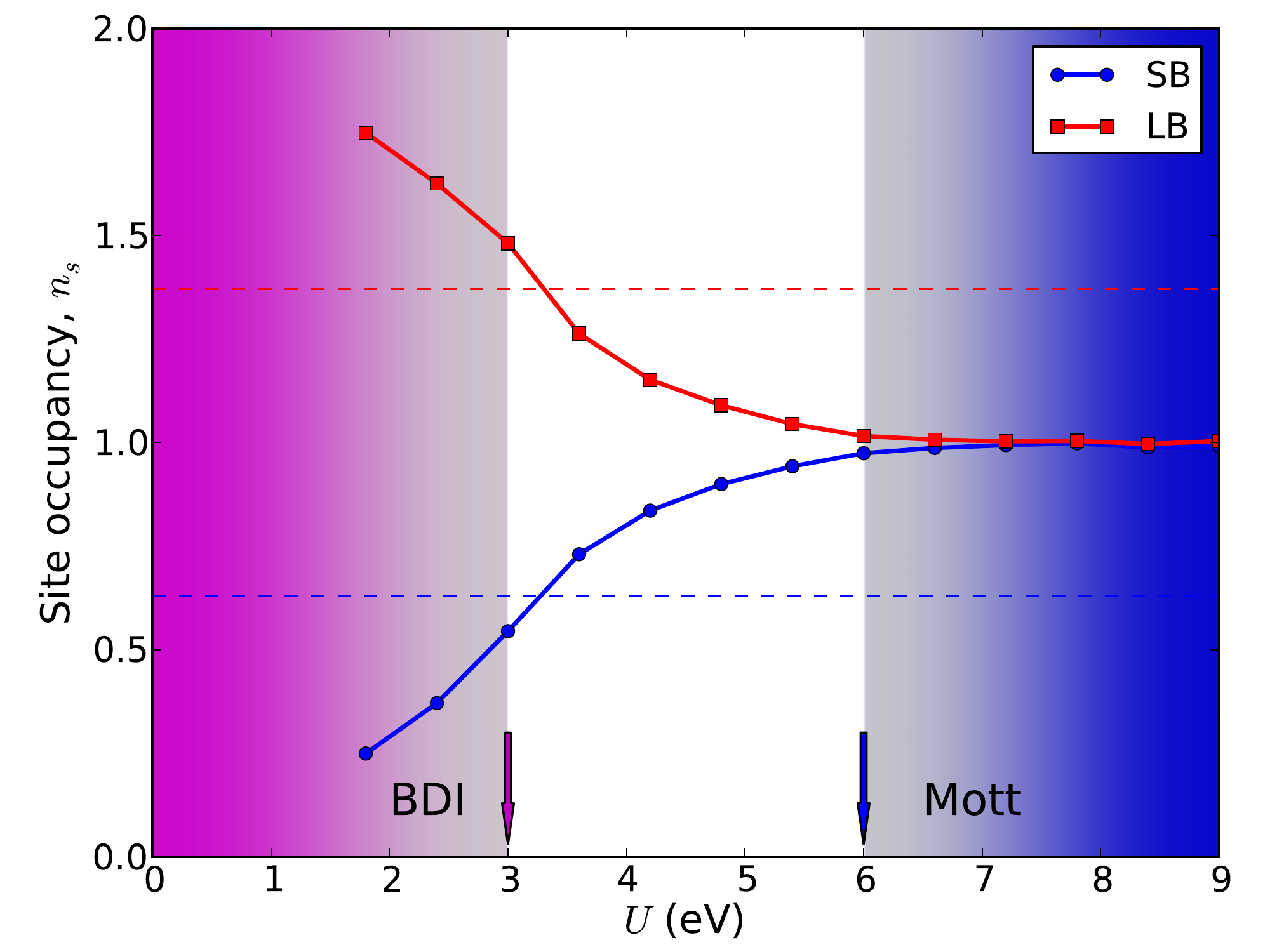}\\
  \caption{(Color online) Occupancies of LB and SB sites as a function
    of $U$ ($J = 0.9$ eV) for a two-sublattice simplified model. The
    magenta (left) and blue (right) arrows indicate the approximate
    positions of the BDI-Metal ($U_{c_1}$) and Metal-Mott ($U_{c_2}$)
    transitions, respectively. The dashed horizontal lines correspond
    to the values of LB and SB occupancies in the non-interacting
    model.}
  \label{fig:ndisp_vs_u}
\end{figure}

This simple estimate provides us with an important insight into the
peculiar behavior of the model for $U < \delsite$. The main conclusion is
that the phase diagram (see the top left panel of Fig.~\ref{fig:model})
of such a system is characterized by 
two transitions Metal $\to$ BDI  and BDI$\to$ Metal at $\Ucs$ and $\Ucl$, respectively,
provided that the site splitting $\delsite$ is larger that some critical
value $\delc$. 
As an illustrative example, the evolution of the site occupancy with increasing $U$ 
for a model calculation (as described in Section~\ref{sec:discussion})is demonstrated
in Fig.~\ref{fig:ndisp_vs_u}. For smaller $\delsite$ the system will remain metallic
at all values of $U < \delsite$.
%

The above considerations can now be extended to $J > 0$.
Now, in the large-$\delsite$ limit the two electrons on LB-sites will 
occupy different orbitals, with the electron spins being aligned on one site and
with the LB-sublattice being antiferromagnetically ordered.
The effective hopping in this case is estimated as
\begin{equation}
\teff = t' + \frac{t^{2}}{\delsite - (U - 3 J)},
\label{eq:teff_j}
\end{equation}
and, importantly, the critical value of $U$ is determined by the relation
$U_{c} + J = \alpha \Weff$ appropriate for a half-filled band.
The equation for $U_{c}$ now reads
\begin{align}
\tilde{U}_{c}^{2} - [(\delsite - 4J) + 4\alpha t'] \tilde{U}_{c} +
\alpha 4 t^{2} + \delsite(4J - t') = 0,
\end{align}
where $\tilde{U} = U - 3J$, with the two solutions
\begin{align}
\tilde{U}_{c} = & \frac{1}{2}\left([\delsite - (4 J - W')] \pm
\sqrt{(4 J - W' + \delsite)^{2} - W^{2}}\right), \label{eq:uc_j}\\
\Ucs \simeq & \,W' - J + \frac{W^{2}}{4 \delsite}, \\
\Ucl \simeq &\, \delsite + 3J - \frac{W^{2}}{4 \delsite},
\label{eq:uc2_j}
\end{align}
provided that $\delsite > \delc = W + W' - 4 J$.
The upper asymptotic boundary is now shifted by $3 J$ compared to the case
wtih $J = 0$ and it is defined by the equation $U - 3J - \delsite = 0$.
As to the lower critical value $\Ucs$ it can now easily become
negative implying that the metallic phase is unstable even for
small values of $U$ (note, also, that the two-sublattice model becomes
generally unstable whth respect to even stronger charge disproportionation
for $U < J$, as mentioned in the main text).

This solution reveals the drastic effect of $J$ on the behavior of the system. Indeed, 
the two roots in (\ref{eq:uc2_j}) are real provided: $\delsite\geq\delc$, with 
$\delc=W+W^\prime-4J$. One sees that the critical disproportionation is now dramatically reduced
by $4 J$. As a result, there is a critical value of the Hund's coupling,
$J_{c} = (W + W') / 4$, above which the system is expected to disproportionate 
spontaneously even at $\delsite=0$. The phase diagram
for this case is sketched in the top right panel of Fig.~\ref{fig:model}.

\bibliography{refs}

\end{document}